\documentclass[times,12pt]{article}
\usepackage[margin=1in]{geometry}
\usepackage[hidelinks]{hyperref}
\usepackage{lineno}
\usepackage{amsmath, amssymb}
\usepackage{hyperref}
\usepackage{graphicx}
\usepackage[round, comma]{natbib}
\usepackage{psfrag}
\usepackage[lofdepth,lotdepth]{subfig}
\usepackage{mathtools}
\usepackage{multirow, authblk}
\usepackage{multicol}
\usepackage{bm}
\usepackage{threeparttable, tablefootnote}
\usepackage{xcolor}
\usepackage{soul}
\extrafloats{100}
\usepackage{subfig}
\newcommand{\GG}[1]{}
\hypersetup{
    colorlinks,
    linkcolor={red!50!black},
    citecolor={blue!50!black},
    urlcolor={blue!80!black}
}

\begin{document}

\begin{center}
    \textbf{A tenuous, collisional atmosphere on Callisto}
\end{center}

\noindent Shane R. Carberry Mogan$^{a, b, c, *}$, Orenthal J. Tucker$^d$, Robert E. Johnson$^{e, f}$, Audrey Vorburger$^g$, Andre Galli$^g$, Benoit Marchand$^h$, Angelo Tafuni$^i$, Sunil Kumar$^{c}$, Iskender Sahin$^a$, Katepalli R. Sreenivasan$^{a, b, f}$\\

\noindent $^a$Tandon School of Engineering, New York University, New York, USA; $^b$Center for Space Science, New York University Abu Dhabi, Abu Dhabi, UAE; $^c$Division of Engineering, New York University Abu Dhabi, Abu Dhabi, UAE; $^d$NASA Goddard Space Flight Center, Maryland, USA; $^e$University of Virginia, Virginia, USA; $^f$Physics Department, New York University, New York, USA; $^g$Physikalisches Institut, University of Bern, Bern, Switzerland; $^h$High Performance and Research Computing Center, New York University Abu Dhabi, UAE; $^i$School of Applied Engineering and Technology, New Jersey Institute of Technology, New Jersey, USA 

\begin{abstract}

A simulation tool which utilizes parallel processing is developed to describe molecular kinetics in 2D, single- and multi-component atmospheres on Callisto. This expands on our previous study on the role of collisions in 1D atmospheres on Callisto composed of radiolytic products \citep{carberrymogan2020} by implementing a temperature gradient from noon to midnight across Callisto's surface and introducing sublimated water vapor. We compare single-species, ballistic and collisional O$_2$, H$_2$ and H$_2$O atmospheres, as well as an O$_2$+H$_2$O atmosphere to 3-species atmospheres which contain H$_2$ in varying amounts. Because the H$_2$O vapor pressure is extremely sensitive to the surface temperatures, the density drops several orders of magnitude with increasing distance from the subsolar point, and the flow transitions from collisional to ballistic accordingly. In an O$_2$+H$_2$O atmosphere the local temperatures are determined by H$_2$O near the subsolar point and transition with increasing distance from the subsolar point to being determined by O$_2$. When radiolytically produced H$_2$ is not negligible in O$_2$+H$_2$O+H$_2$ atmospheres, this much lighter molecule, with a scale height roughly an order of magnitude larger than that for the heavier species, can cool the local temperatures via collisions. In addition, if the H$_2$ component is dense enough, particles originating on the day-side and precipitating into the night-side atmosphere deposit energy via collisions, which in turn heats the local atmosphere relative to the surface temperature. Moreover, the difference between H$_2$ atmospheric escape rates in single-species and multi-species atmospheres is small: the H$_2$ only has to diffuse through a few hundred km of the heavier gases before it is the lone species in the atmosphere out to the Hill sphere. Finally, we discuss the potential implications of this study on the presence of H$_2$ in Callisto's atmosphere and how the simulated densities correlate with expected detection thresholds at flyby altitudes of the proposed JUpiter ICy moons Explorer (JUICE) spacecraft.

\end{abstract}

\noindent *Corresponding author at: Center for Space Science, New York University Abu Dhabi, Abu Dhabi, UAE. \\
\noindent \textit{E-mail address}: ShaneRCM@nyu.edu (Shane R. Carberry Mogan).

\newpage
\section{Introduction}

Understanding the composition and behavior of tenuous atmospheres can give important information on their evolution on small solar system bodies. The atmosphere on Callisto, which is the focus of this study, is very poorly constrained, and its origin and evolution are not well understood. A tenuous CO$_2$ atmosphere was observed by \textit{Galileo}, and its extent was suggested to be global due to the volatility and mobility of CO$_2$ at Callisto's surface temperatures \citep{carlson1999}. \textit{Galileo} radio occultations indicated the presence of a transient but substantial ionosphere, which was suggested to be sourced by a collisional O$_2$ atmosphere \citep{kliore2002}, $\sim$2 orders of magnitude more dense than that of the observed of CO$_2$. Atomic oxygen emissions detected using the Hubble Space Telescope (HST)-Cosmic Origins Spectrograph were suggested to be induced by photoelectron impacts in a O$_2$-dominated atmosphere \citep{cunningham2015}, with a derived O$_2$ column density an order of magnitude less than that suggested by \cite{kliore2002}. However, even this reduced estimate for the thickness of Callisto's atmosphere is only exceeded among other solar system satellites by those on Io, Triton, and Titan. In addition, the denser O$_2$ atmosphere inferred by \cite{kliore2002} was derived when the trailing hemisphere was sunlit, whereas the estimates of \cite{cunningham2015} were derived when the leading hemisphere was sunlit. Finally, faint atmospheric emissions above Callisto's limb were recently detected from HST-Space Telescope Imaging Spectrograph observations \citep{roth2017}, likely originating from resonant scattering by an H corona, which was suggested to be produced via dissociation of H$_2$O and H$_2$, with much smaller amounts due to sputtering and radiolysis of its icy surface. We refer the reader to \cite{carberrymogan2020}, hereafter referred to as C20, for a more thorough discussion of the observations and previous models of Callisto's atmosphere.

Since Callisto's atmosphere transitions from a relatively thin but collisional atmosphere to free molecular flow above the exobase, it cannot be accurately described by solving fluid equations (e.g., Navier-Stokes) which break down when the length scales associated with gradients in the flow field become comparable to the mean free path. Moreover, recent ballistic models of Callisto's atmosphere (e.g., \citealt{vorburger2015, vorburger2019}) can fail to capture the influence collisions have on the structure and escape. Thus, in order to accurately model the transition from collisional to collisionless flow in Callisto's atmosphere, a molecular kinetics model is required. In an earlier paper (C20), we used 1D, spherically symmetric molecular kinetics models to demonstrate the influence of collisions and thermal escape in atmospheres on Callisto composed of radiolytic products which thermally desorb from and, on return, permeate the porous regolith. We showed that, at densities suggested by observations and models, collisions played a critical role in determining how the diffusion of a light species through heavier background gases and its escape affected the evolution and structure. Here we extend that work to 2D by including an angular axis (subsolar latitude) in order to demonstrate the effects on the spatial distribution of species desorbing or sublimating from a surface that is non-uniformly illuminated. We also varied the density of the yet to be detected light species, H$_2$, over orders of magnitude to constrain the possible structure of Callisto's atmosphere in preparation for the expected spacecraft studies of the Jovian system by ESA's JUpiter ICy moons Explorer (JUICE).

Following our earlier work (C20), using parallel processing, we developed a flexible simulation tool to simulate molecular kinetics in multi-dimensional, multi-species atmospheres that range from highly collisional near the surface to free molecular flow above the exobase. This software tool, described in Appendix \ref{parallel}, is applied to several 2D model atmospheres on Callisto composed of O$_2$, H$_2$, and H$_2$O components using source rates suggested by previous modeling efforts for the O$_2$ and H$_2$O and a range of rates for the H$_2$. Since the emphasis here is on the roles that H$_2$ and H$_2$O play, we do not include the observed CO$_2$ component as this would simply add to the relatively heavy background gases (e.g., O$_2$) close to the surface. The species comprising these simulated atmospheres have very different spatial distributions, as their local source rates are driven by the temperature gradient across the surface, something our previous 1D models could not capture. Unfortunately, none of the densities for these components are well constrained. The H$_2$O, which, as discussed below, has been suggested to dominate near the subsolar point, and the radiolytic product H$_2$, which is produced concomitantly with O$_2$, are both expected to be present but neither species has been directly detected. Therefore, we implement these model atmospheres and discuss interactions between the various species as a guide to understanding how the atmosphere would be affected by the variation of the surface temperature with subsolar latitude and changes in the source rates.

Since Callisto is a relatively low density outer solar system object, it is suggested to have a significant solid H$_2$O component. Further, it has the warmest surface temperature and the lowest albedo among the icy Galilean satellites (\citealt{moore2004} and references therein). Thus, when the surface is illuminated and heats up to its relatively large day-side temperatures, sublimation of water ice dominates surface processes, as described below. Understanding the fate of the sublimated water vapor can lead to a better understanding of the evolution of Callisto's atmosphere. We briefly discuss observations and corresponding experiments and models pertaining to H$_2$O at Callisto below.

Early high resolution spectral studies failed to positively identify water frost on Callisto but were used to set an upper limit of 15$\pm$10$\%$ for its surficial coverage \citep{pilcher1972}. A few years later the first positive evidence for water frost on Callisto's surface was detected \citep{lebofsky1977}. Based on reflectance spectra of Callisto's leading and trailing hemispheres, \cite{pollack1978} suggested that substantial quantities of water bound to darker materials are present in the surface but there was relatively little evidence for ice on the surface. Moreover, since Callisto has the lowest albedo among the Galilean satellites, they suggested the surface is dominated by refractory materials, such as carbonaceous chondrites. However, subsequent spectral reflectance measurements made by \cite{clark1980galilean} and \cite{clark1980ganymede} found no evidence for water molecules bound to dark material, and they estimated that only an upper limit of 5$\pm$5$\%$ may be present. \cite{clark1980galilean} suggested water appears in the form of ice patches and the areal coverage of water frost is $\sim$20--30$\%$ on Callisto's leading hemisphere. \cite{clark1980ganymede} suggested even more water ice could be present on the surface, $\sim$30--90 by weight $\%$ (wt$\%$), but that non-ice materials are also present and the ice and non-ice materials are mixed to some extent. \cite{spencer1987a} later reevaluated these spectra and concluded they were consistent with a segregated (ice and non-ice) surface; they suggested that the bulk of the surface was covered in ice-free, carbonaceous chondrite-like material containing some bound water molecules, consistent with \cite{pollack1978}, with only $\sim$10$\%$ areal coverage by ice. \cite{spencer1987b} suggested this segregation of Callisto's surface materials to be a result of ice sublimating from a darker, warmer region and preferentially depositing onto brighter, cooler icy patches. This process, referred to as cold-trapping, continues until a lag deposit develops on the now ice-free, dark region and cuts off sublimation. \cite{roush1990} compared laboratory spectral reflectance measurements of homogeneous and heterogeneous mixtures of ice and non-ice components to the reflectance spectra discussed above. Their best comparisons yielded 12--14$\%$ and 14--16$\%$ areal coverage of water ice and 37--43 wt$\%$ and 26--37 wt$\%$ on the leading and trailing hemispheres, respectively. Subsequent extensive modeling of ice and non-ice spectra by \cite{calvin1991} yielded similar results (20-45 wt$\%$), and Callisto's surface was suggested to consist of patches of mixed ice and rocky materials.

Images of Callisto's surface taken by \textit{Voyager 1} and \textit{2} revealed it to be covered in large craters, possibly dating back $\sim$4 billion years \citep{smith1979a, smith1979b}. Whereas \textit{Voyager} images showed a polar frost was present on Ganymede, no such phenomena was observed on Callisto \citep{squyres1980}. \cite{spencer1987b} suggested this apparent lack of poleward migration of ice was indirect evidence for local segregation of ice on Callisto's surface. Images taken by \textit{Voyager 1} of craters in Callisto's north polar region revealed an anomalous appearance: whereas bright volatile deposits, likely water ice, are cold-trapped on the north-facing slopes with lower mean solar illumination angles and hence lower surface temperatures, the slopes facing the sun appeared darker \citep{spencer1984}. Thus, \cite{spencer1984} suggested that thermal sublimation controlled by contrasts in the local insolation-dependent surface temperature dominates ice migration at high latitudes, and that other migration mechanisms, such as sputtering or micrometeorite bombardment, likely only play a minor role. \cite{moore1999} suggested a similar process can lead to the significant mass wasting and landform degradation observed on Callisto in high resolution images taken by \textit{Galileo}.

%High resolution images taken by \textit{Galileo} of Callisto's surface revealed it to be the most degraded surface of the Galilean satellites with a surprisingly few small craters, implying that degradation processes were ongoing \citep{moore1999}. Consistent with earlier observations, Callisto's surface was also observed to be segregated, either bright or dark with little variation in between. Whereas the former feature is likely due to water frost near or at the crests of high standing topography, such as crater rims, the latter feature is almost always present in low-lying areas, such as within crater floors. \cite{moore1999} demonstrated that sublimation-driven landform modification and mass wasting (i.e., ``sublimation-degradation'') could be responsible for these features. That is, dark material at the base of a scarp within a crater traps heat and preferentially warms nearby ice. The subsequent sublimation of the ice at the scarp base maintains the scarp's steepness while causing it to retreat. As the dark material is left behind and accumulates at the base of the slope, the scarp continues to retreat while the sublimated ice accumulates as frost on the crest of the crater. Eventually, oversteepening of the scarp causes mass wasting. Consistent with interpretations of \textit{Voyager} images (e.g., \citealt{spencer1984}), degradation of Callisto's surface was suggested to be dominated by sublimation as opposed to other processes, such as sputter ablation and/or impact erosion.

As mentioned above, HST observations of Callisto detected faint atmospheric emissions above its limb, likely originating from resonant scattering by a H corona \citep{roth2017}. Atomic hydrogen can be produced at Callisto via dissociation of hydrogen bearing molecules in its atmosphere, such as H$_2$O and H$_2$, with much smaller amounts due to sputtering and radiolysis of its icy surface. The resulting H corona was observed to be denser when observing Callisto's leading hemisphere at eastern elongation than when observing its trailing hemisphere at western elongation, opposite the O$_2$ asymmetry described earlier. \cite{roth2017} suggested this asymmetry could originate from differences in sublimation of the surface ice as Callisto's leading hemisphere is darker than its trailing hemisphere (\citealt{moore2004} and references therein). However, these authors cautioned that the lower albedo on the leading hemisphere due to increased surface coverage by darker non-ice surface material, might not imply a higher H$_2$O source rate. Moreover, the lower H Ly-$\alpha$ signal in the observation of the trailing hemisphere was later shown to be possibly caused by absorption in the Earth geocorona \citep{alday2017}. Thus the hemispheric asymmetry might be due to differences in source rates or the Earth geocorona effect. Analogous to models at Ganymede \citep{marconi2007}, near the subsolar point, where the surface temperatures and, hence, sublimation rates are presumably the highest, \cite{roth2017} suggested dissociation of water vapor is likely the primary source for atmospheric H, whereas beyond the terminator H$_2$ produced by sputtering/radiolysis might dominate.

Building on similar 1D studies carried out in C20, here we quantify the effect of intermolecular collisions and diurnal transport on thermal escape and the temperature and compositional structure of Callisto's atmosphere. In Section \ref{Sec2}, we describe how we model these atmospheres using the direct simulation Monte Carlo method \citep{bird1994}, which is often used to simulate tenuous atmospheres. In Sections \ref{Sec3} and \ref{Sec4} we present and discuss the implications of our results as well as various parameters that will be considered in subsequent studies, and conclude this study in Section \ref{Sec5}.

\section{Numerical Method} \label{Sec2}

The direct simulation Monte Carlo (DSMC) method \citep{bird1994} is used here to simulate the thermal evolution of 2D, single- and multi-component atmospheres on Callisto. DSMC is a widely used computational method for simulating rarefied gas dynamics, such as flow transitioning from local thermal equilibrium (LTE) to nonequilibrium. It is used to simulate macroscopic gas dynamics by directly modeling stochastic microscopic processes of individual molecules in the gas. Molecular kinetics is simulated via computational particles, each of which represents a large number of real atoms or molecules and is characterized by position, velocity, and internal state. As these particles traverse physical space they are influenced by gravitational forces and binary collisions.

A DSMC simulation is discretized into time-steps, $\Delta t$, which are much smaller than the mean time between collisions, allowing particle motion to be decoupled from particle collisions. The computational domain which spans physical space is discretized into a grid comprised of cells of varying extent. \cite{bird1994} suggests the extent of a cell should be a function of the local mean free path (MFP), $\ell_\mathrm{MFP} = v_{th} / \nu_\mathrm{coll}$, where $v_{th} = \sqrt{\frac{8 k_B T}{\pi m}}$ is the mean thermal speed and $\nu_\mathrm{coll}$ is the collision frequency, which is proportional to the local number density, $n$, and collision cross-section, $\sigma$. Here $k_B$ = 1.38$ \times 10^{-23}$ m$^2$ kg s$^{-2}$ K$^{-1}$ is the Boltzmann constant, $T$ is temperature, and $m$ is the mass of the atmospheric species. This network of cells allows for particles to be locally grouped together based on their position in order to compute collisions. If the extent of a cell is much larger than $\ell_\mathrm{MFP}$, the particles separated by such a distance could collide, allowing a non-physical transfer of energy to occur. To resolve this issue, ``sub-cells'' can be implemented. That is, for the purpose of computing collisions the original cell can be divided into smaller cells. Particles within the original cell are then re-grouped into this smaller subset of cells and, within these sub-cells, collision partners are selected and collisions are computed. Thus, even in a cell whose extent is larger than $\ell_\mathrm{MFP}$, collisions are ensured to occur only between particles within a physically realistic distance ($\leq \ell_\mathrm{MFP}$) from one another. Collisions are statistically computed between randomly selected particles in a cell or sub-cell based on their relative velocities and $\sigma$. Using the variable hard sphere (VHS) model \citep{bird1994}, $\sigma$ is calculated based on the relative speed of the particles and the molecular diameter, $d$, and the viscosity index, $\omega$, of the species; for inter-species collisions, $d$ and $\omega$ are calculated as averages of the two species' parameters. Here we simulate collisions between particles, where the redistribution of energy among kinetic (translational) and internal (rotational) modes is calculated using the Larsen-Borgnakke (LB) model \citep{larsen1974}.

The initial state of a computational domain can be a vacuum or initialized to an assumed state (e.g., a hydrostatic isothermal atmosphere). Particles are injected into the domain from the surface with velocities and internal states determined here by the local surface temperature (e.g., velocities sampled from a Maxwellian flux distribution \citep{smith1978}). As they incrementally move throughout the domain in distances equivalent to the product of their velocity and $\Delta t$, they can change cells, as well as interact with the boundaries. Here particles absorbed at the surface or that escape to space are discarded and their allocated computational memory is ``recycled'' for new particles entering the domain at a later time-step. Each particle is attributed a ``weight,'' which defines the number of real molecules it represents and is a function of its source rate; see Appendix \ref{partweight} for how particle weights are calculated here.

Finally, a steady-state DSMC simulation is run long enough such that the distribution of particles and their corresponding characteristics yield steady-state macroscopic properties, such as density and temperature. Steady-state is determined through periodic sampling of the flow field. For example, when the flow field is sampled, the velocity vectors of all particles in each cell relative to a local mass averaged velocity vector are tabulated and the magnitude is squared and summed to determine the average squared thermal speed ($\overline{c'^2}$), which is then used to calculate the translational temperature ($T$): $\frac{3}{2} k_B T = \frac{1}{2} m \overline{c'^2}$. Furthermore, the average radial and tangential velocities ($v_r, v_\phi$) among particles in each cell are then used to determine the corresponding temperatures ($T_r, T_\phi$). As demonstrated in C20, when the flow is collisional and in LTE, $T_r$, $T_\phi$, and $T$ are equal, and when flow transitions to nonequilibrium, they will diverge from one another. Since there is a relatively small number of particles relative to the large number of real atoms or molecules they represent, there will be significant noise in an instantaneous sample. Therefore, successive samples are carried out in order to increase the sample size and thus reduce the statistical noise from the results. Increasing the number of particles in a simulation can also reduce the statistical noise; however, doing so comes at a computational cost as it increases the number of particle-based calculations made at each time-step.

\subsection{2D Model} \label{2Dmodel}

Although our 1D model in C20 was sufficient to demonstrate the importance of intermolecular collisions in Callisto's atmosphere, a multi-dimensional model is required to simulate the day-to-night temperature gradient across the surface. Here 2D simulations are carried out for both single- and multi-component atmospheres. In a 2D spherical domain, the grid is decomposed into cells that vary radially as well as along the subsolar latitude (SSL), from noon (SSL$=0^\circ$) to midnight (SSL$=180^\circ$) assuming that the domain is symmetric about the axis passing through these points. Similar 2D, axisymmetric DSMC models have been applied to other Galilean satellites, such as Ganymede \citep{marconi2007} and Io \citep{austin2000, zhang2003, zhang2004}, as well as to cometary comae \citep{combi1996}. The extent of our angular cells have uniform increments of $\Delta$SSL$= 1^\circ$ from SSL$=0^\circ$ to SSL$=180^\circ$ or, in the case of the single-species H$_2$O atmosphere, SSL$=90^\circ$. We vary the temperature across Callisto's surface from the subsolar point (noon) to the anti-solar point (midnight) using the model from \cite{hartkorn2017}

\begin{equation}
T(\mathrm{SSL}) = \frac{1}{2} (T_\mathrm{noon}+T_\mathrm{midnight}) + \frac{1}{2} (T_\mathrm{noon}-T_\mathrm{midnight}) \cos(\mathrm{SSL}),
\end{equation} \label{TempEq}

\noindent where $T_\mathrm{noon} = 155$ K and $T_\mathrm{midnight} = 80$ K are the approximate noon and midnight temperatures, which are consistent with observed mean subsolar and predawn equatorial nighttime temperatures (\citealt{moore2004} and references therein). Radial cells are generated from Callisto's surface ($r_C = 2410$ km), up to a predefined upper spherical boundary ($r_\mathrm{max}$). See Appendix \ref{2dgrid} for more specifics about how the radial extent of cells are determined for the various atmospheres and how that correlates with the angular extent of cells. For single- and multi-component O$_2$ and H$_2$O atmospheres, which have negligible escape rates, $r_\mathrm{max}$ is set to $r_\mathrm{max}-r_C \sim 1,200$ km, and for atmospheres in which H$_2$ is present, $r_\mathrm{max}$ is set to Callisto's Hill sphere radius, $r_\mathrm{HS} = \left( \frac{M_C}{3 M_J} \right)^{1/3} d_{JC} \sim 20.8 r_C$ ($r_\mathrm{max}-r_C \sim 48,000$ km). Here $M_C$ = 1.08$ \times 10^{23}$ kg and $M_J$ = 1.898$ \times 10^{27}$ kg are the masses of Callisto and Jupiter, respectively, and $d_{JC} \sim 26.3 r_J$ is the distance from Callisto to Jupiter in units of Jupiter radii, $r_J = 71,492$ km.

Since we choose to use upward particle fluxes at the lower boundary, at the beginning of the simulation particles are injected into initially empty cells across the surface with velocities sampled from a Maxwellian flux distribution based on the local surface temperature. As particles traverse physical space, their movement is tracked in 3D Cartesian coordinates using a 4th-order Runge Kutta integration. After each $\Delta t$ their new radial position and SSL coordinate are calculated to check if they leave their current cell or interact with the boundaries. Particles that return to the lower boundary of the domain are assumed to enter and permeate the porous regolith and their allocated computational memory is recycled. In single- and multi-component O$_2$ and H$_2$O atmospheres, in the rare case that a particle exceeds $r_\mathrm{max}$ it is tracked along its ballistic trajectory until it eventually returns to the domain. When H$_2$ is present and $r_\mathrm{max} \sim r_\mathrm{HS}$, any H$_2$ particle whose displacement exceeds $r_\mathrm{max}$ is assumed to have escaped and is removed from the simulation and its allocated computational memory is recycled.

\subsection{Surface Fluxes} \label{surfflux}

The surface flux for the radiolytic product O$_2$, for which escape is negligible, is estimated from the local temperature and the same uniform, surface density as in C20 ($n_{0, \mathrm{O_2}}$ $\sim$ 8$ \times 10^8$ cm$^{-3}$). This is about $2 \times$ that of the observed CO$_2$, roughly consistent with the column densities suggested in \cite{hartkorn2017}: $\sim$1--3$\times 10^{15}$ cm$^{-2}$ with an O$_2$ scale height at 155 K of $\sim$30 km. If condensed on the surface, such a column is equivalent to only about two monolayers of O$_2$.

The H$_2$ component, which can affect Callisto's atmospheric structure (e.g., C20), is a radiolytic and photolytic product from the surface ice and the expected water vapor (e.g., \citealt{liang2005, marconi2007, vorburger2018}), as well as from the dark meteoritic lag deposit (e.g., \citealt{johnson1991}) and even proton implantation (e.g., \citealt{tucker2021}). Due to the lack of observational constraints, we first ignore the H$_2$ and then vary it's surface flux over orders of magnitude. The upper limit was estimated by assuming the very porous surface regolith is permeated with a stochiometric mix of adsorbed/trapped gaseous products. Although H$_2$ has a small escape rate and O$_2$ is likely reactive in the regolith, without further constraints the maximum flux from the porous regolith was assumed to have an approximate 2:1 H$_2$:O$_2$ ratio. The intermediate and lowest fluxes applied here are 0.1 and 0.01 times that of C20, respectively. The latter is roughly consistent with ignoring the presence of absorbed/trapped volatiles in the regolith and assuming that, once produced, the radiolytic H$_2$ and O$_2$ remain in the atmosphere subject only to escape and gas phase destruction processes, as in simulations at the companion satellites with predominantly ice covered surfaces (e.g., \citealt{marconi2007, leblanc2017, vorburger2018}). Even smaller fluxes are, of course, possible, in which case the results are similar to those in which H$_2$ is not included.

As is the case for H$_2$, gas phase H$_2$O has not been directly observed. However, the bright areas, which are only a fraction of the surface (e.g., \citealt{spencer1987a}), are assumed to be dominated by ice. Therefore, gas-phase H$_2$O is assumed to be present at some level as discussed earlier. Here we assume that the global sublimation of H$_2$O is determined by Callisto's relatively warm surface temperatures, as in other models (e.g., \citealt{liang2005, vorburger2015, hartkorn2017}). Since the sublimation rate is extremely sensitive to the diurnal variation of Callisto's surface temperature, $n_{0, \mathrm{H_2O}}$ varies by several orders of magnitude across the surface. To approximate this, we use a global, average concentration of ice on the surface ($c_\mathrm{H2O}$) to calculate the density variation with SSL estimated from the vapor pressure ($P_v$) used in \cite{vorburger2015}

\begin{equation}
\ln \left( \frac{P_v(T_0(\mathrm{SSL}))}{P_t} \right) = \frac{3}{2} \ln \left( \frac{T_0(\mathrm{SSL})}{T_t} \right) + \left( 1 - \frac{T_t}{T_0(\mathrm{SSL})} \right) \eta \left( \frac{T_0(\mathrm{SSL})}{T_t} \right),
\end{equation}

\noindent where $P_t$ = 6.1166$ \times 10^{-3}$ bar and $T_t$ = 273 K are the triple point pressure and temperature for water, respectively, and $\eta$ is a polynomial relation between the surface temperature, $T_0$, and $T_t$. Using a concentration $c_\mathrm{H2O}$ = 10$\%$ and the Ideal Gas equation, $n_{0, \mathrm{H_2O}} (\mathrm{SSL}) = \frac{c_\mathrm{H2O} P_v(T_0(\mathrm{SSL}))}{k_B T_0(\mathrm{SSL})}$, we obtain a surface density at the subsolar point, $\sim$10$^9$ cm$^{-3}$, and distribution along the SSL consistent with the literature (e.g., \citealt{liang2005, vorburger2015, hartkorn2017}). 

\begin{figure}[h!]
    \centering
    \includegraphics[scale=0.28]{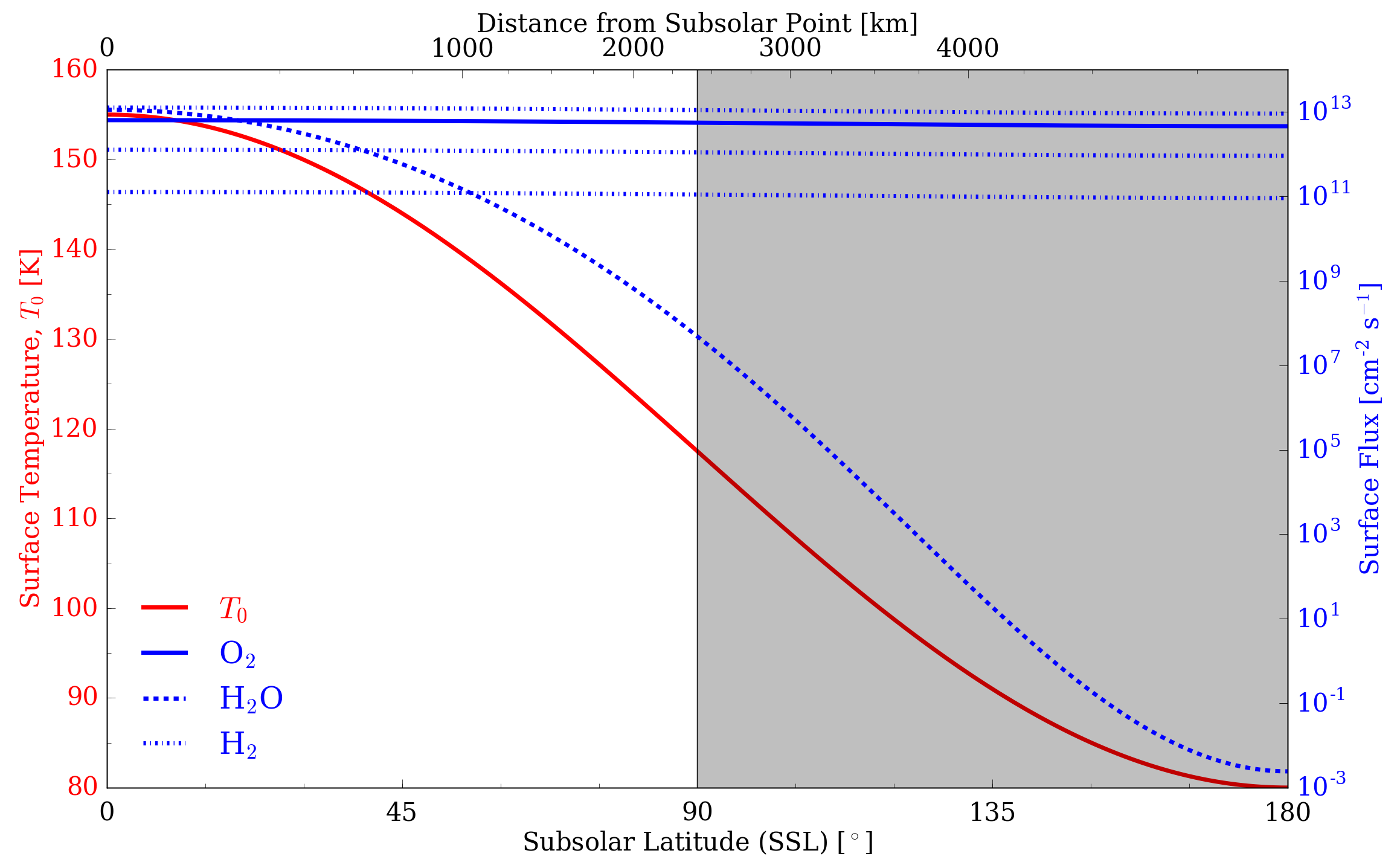}
    \caption{Temperature distribution (solid red line) and surface fluxes for H$_2$O, H$_2$, and O$_2$ (solid, dashed, and dashed-dotted blue lines, respectively) across Callisto's surface from the subsolar point (SSL=0$^\circ$) to the anti-solar point (SSL=180$^\circ$). The terminator (SSL=90$^\circ$) separates the ``Day-side'' (white background) from the ``Night-side'' (grey background). Note, as described above, we apply 3 different values for H$_2$ surface fluxes, each of which is represented with a dashed blue line and they differ by the magnitude; and the O$_2$ and H$_2$ fluxes vary as $\sqrt{T_0}$ across the surface, $\sim$1.4 from SSL=0$^\circ$ to SSL=180$^\circ$.}
    \label{fig:SourceRates}
\end{figure}

Although the surface fluxes of these 3 species are very poorly constrained, these estimates are applied in simulations to guide our understanding of their potential roles in this important atmosphere, which will be the subject of many forthcoming observations. The surface temperature distribution used, as well as the corresponding surface fluxes of H$_2$O, O$_2$, and H$_2$, are plotted in Fig \ref{fig:SourceRates}.

\section{Results} \label{Sec3}

We simulate molecular kinetics in model 2D single- and multi-component atmospheres on Callisto using the DSMC method and parallel processing. The influence of collisions and the presence and escape of H$_2$ are studied. For details on the collision parameters for each species see Appendix \ref{collparams}. The nominal exobase occurs when the Knudsen number, Kn $\sim 1$. Here Kn is defined as the ratio between $\ell_\mathrm{MFP}(r)$ and the local atmospheric scale height, $H(r) = \frac{k_B T(r) r^2}{G M_C m}$, where $G$ = 6.674$ \times 10^{-11}$ m$^3$ kg$^{-1}$ s$^{-2}$ is the gravitational constant. In a multi-species atmosphere an average exobase is calculated when Kn$_\mathrm{avg}=\frac{ \sum_i \mathrm{Kn}_i n_i }{\sum_i n_i} \sim 1$, where $i$ represents the species.

As described earlier, only the O$_2$ has been inferred from atmospheric observations, but the H$_2$ and H$_2$O are expected to be present. Therefore, single-species simulations were first carried out in 2D for all three species as both collisional and ballistic atmospheres to understand the influence of collisions on the atmospheric structure and transport, as well as to compare to the 1D results in C20 for the H$_2$ component. These results are then compared to multi-species O$_2$+H$_2$O and O$_2$+H$_2$O+H$_2$ atmospheres.

\subsection{O$_2$ and H$_2$O Atmospheres}

\subsubsection{Single-Species} \label{O2H2O_SingleSpecies}

Collisional and ballistic O$_2$ atmospheres were simulated. Steady-state values of density and temperature were reached in $\sim$2.5$ \times 10^4$ seconds in the former and in $<$10$^4$ seconds in the latter. In both cases the horizontal transport is small, so that, in the absence of additional heat sources, collisions have a very small effect on the variation in atmospheric structure across the surface. Thus, both the collisional and ballistic column densities vary from $N_\mathrm{O_2} \sim 3.4 \times 10^{15}$ cm$^{-2}$ near noon to $\sim$1.6$ \times 10^{15}$ cm$^{-2}$ near midnight, consistent with those calculated by \cite{hartkorn2017}, with isothermal temperature profiles at each SSL across the surface as seen from the density and temperature profiles in Figs. \ref{fig:NumDens_O2} and \ref{fig:O2H2O_TransTemps_SS} in Appendix \ref{profiles}. In the collisional atmosphere, the exobase varies from $r-r_C \sim 100$ km near noon to $r-r_C \sim 50$ km near midnight, whereas it is at the surface in a ballistic atmosphere. A similar comparison was made for collisional and ballistic H$_2$O atmospheres. Steady-state was reached in times similar to those for O$_2$, but the density of H$_2$O, a function of local surface vapor pressure, varies significantly across Callisto's surface becoming negligible near the terminator. The density decreases exponentially with altitude and distance from the subsolar point, with a maximum radial column density of 7.6$ \times 10^{15}$ cm$^{-2}$ as seen in Fig. \ref{fig:NumDens_H2O} in Appendix \ref{profiles}.

\begin{figure}[h!]
    \centering
    \includegraphics[scale=0.26]{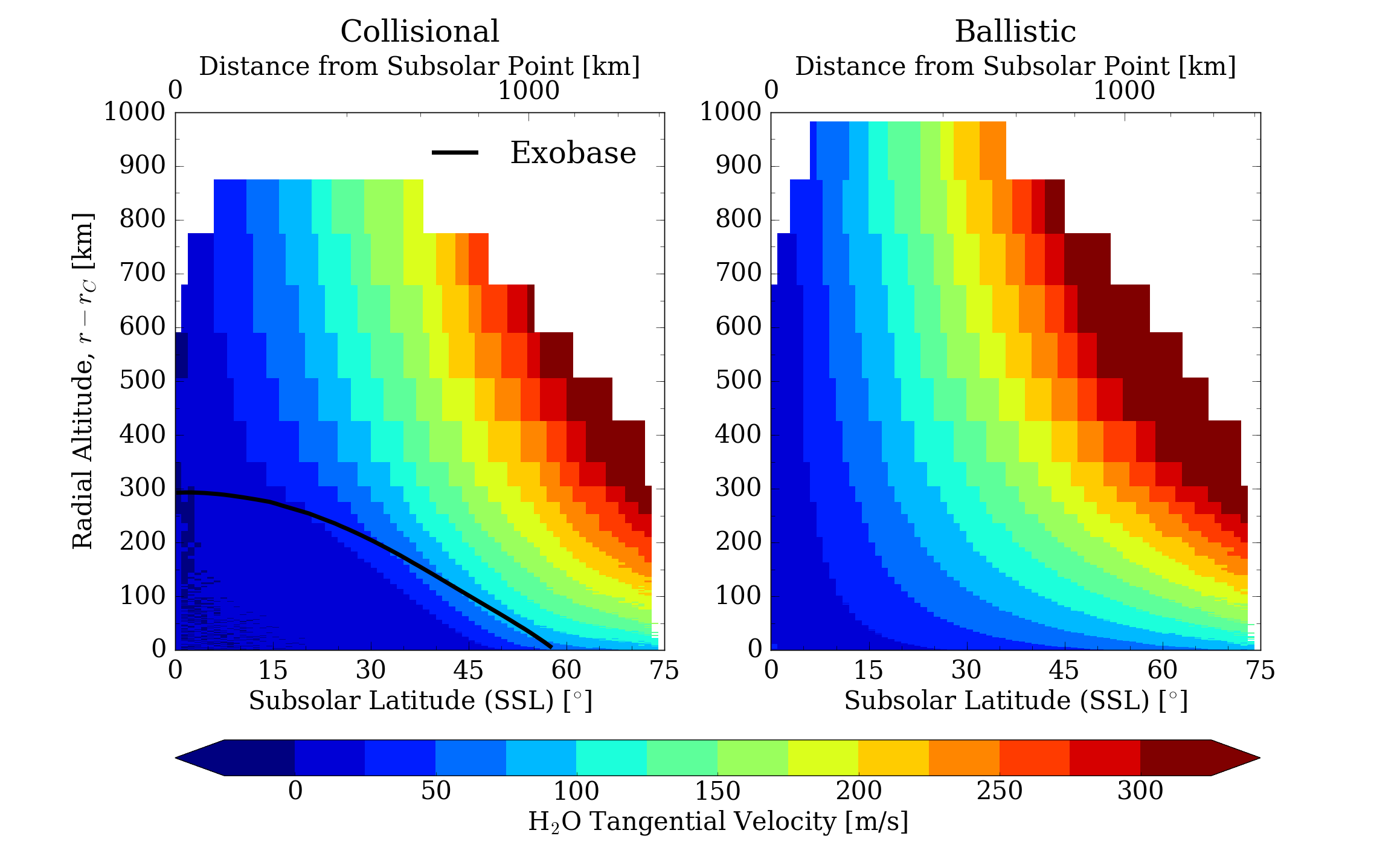}
    \includegraphics[scale=0.26]{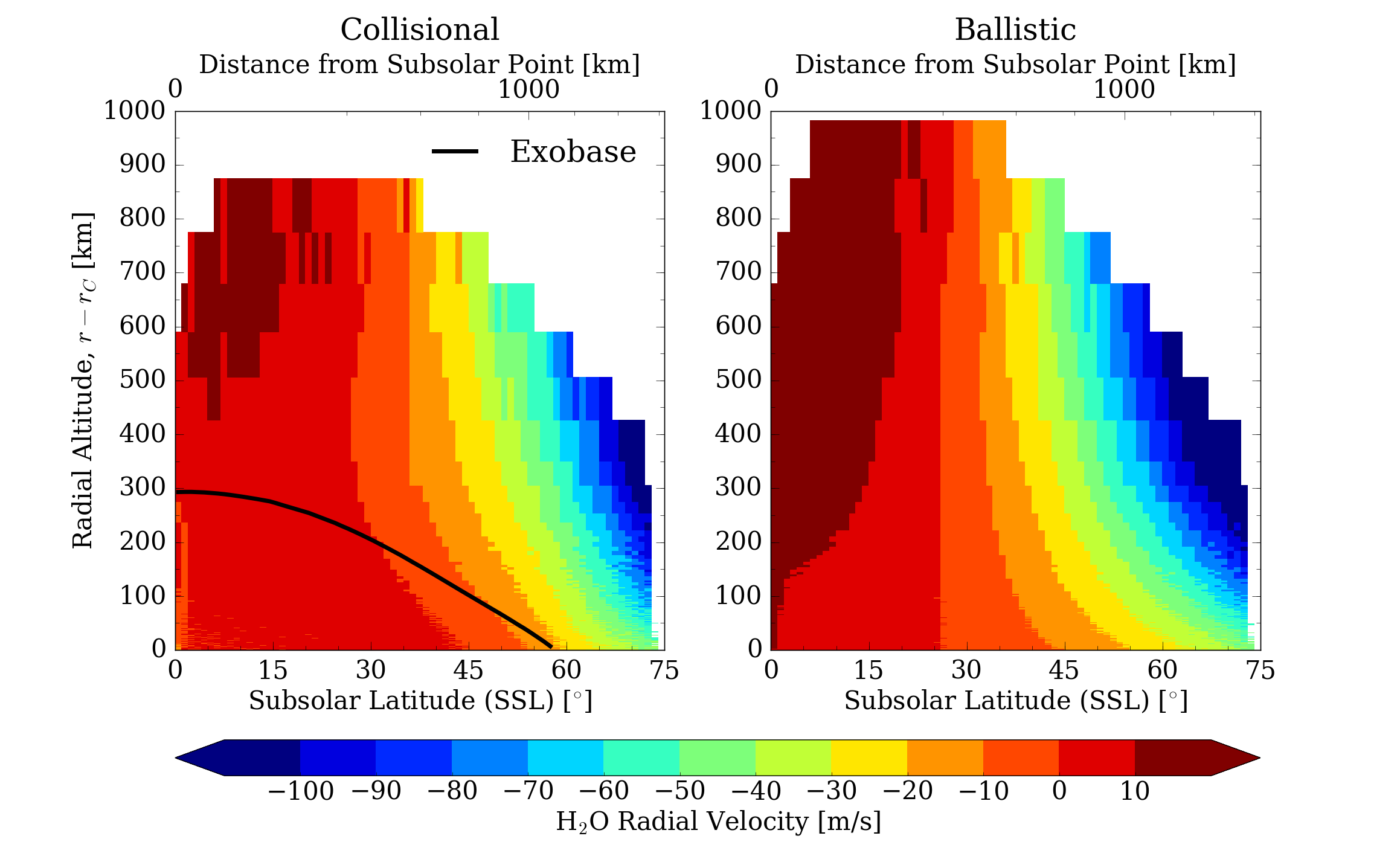}
    \caption{Tangential (\textit{top}) and radial velocities (\textit{bottom}) for single-species, collisional (\textit{left}) and ballistic (\textit{right}) H$_2$O atmospheres vs. subsolar latitude (SSL; x-axis, \textit{bottom}), distance from subsolar point (x-axis, \textit{top}), and radial altitude, $r-r_C$ (y-axis). The exobase is represented by a solid black line in the collisional atmosphere. Note the white regions of the plot represent cells with $<$10 particles.}
    \label{fig:H2O_ZenVel_RadVel}
\end{figure}

Despite the density gradient along the SSL, collisions do not significantly affect the total distance particles travel along the angular axis relative to a ballistic atmosphere so the density profiles are similar. However, collisions convert particles' directional velocities to random motion, resulting in lower tangential and radial velocities below the exobase and out to the more tenuous regions (SSL=30$^\circ$ at $r-r_C \sim 200$ km to SSL=45$^\circ$ at the surface), as seen in Fig. \ref{fig:H2O_ZenVel_RadVel}. Beyond this region, however, the tangential velocities are similar, albeit slightly diminished, relative to those in a ballistic atmosphere, resulting in a similar distribution along the SSL axis. Conversely, collisions diminish the radial extent of an H$_2$O atmosphere due to the difference in the radial velocities, as seen in Fig. \ref{fig:H2O_ZenVel_RadVel}. With increasing distance from the subsolar point, negative radial velocities are generated in the more tenuous regions of the collisional atmosphere as particles fall back to the surface uninhibited by collisions, resembling those in the ballistic atmosphere.

Although the random motion generated by collisions diminishes the tangential and radial velocities relative to a ballistic atmosphere, large temperatures are generated at the outer edges of the distribution, as seen in Fig. \ref{fig:O2H2O_TransTemps_SS} in Appendix \ref{profiles}. A similar phenomenon was observed in models of volcanic plumes on Io \citep{zhang2003}. In regions of the flow field where the density is very low, the gas is a mixture of ballistic particles that have expanded outward along the SSL axis and those that are falling back to the surface. Such a bimodal distribution in a tenuous regime is not Maxwellian and the word ``temperature'' is used as a proxy for describing the mean kinetic energy.

\subsubsection{O$_2$+H$_2$O} \label{O2H2O_MultipleSpecies}

\begin{figure}
    \centering
    \includegraphics[scale=0.22]{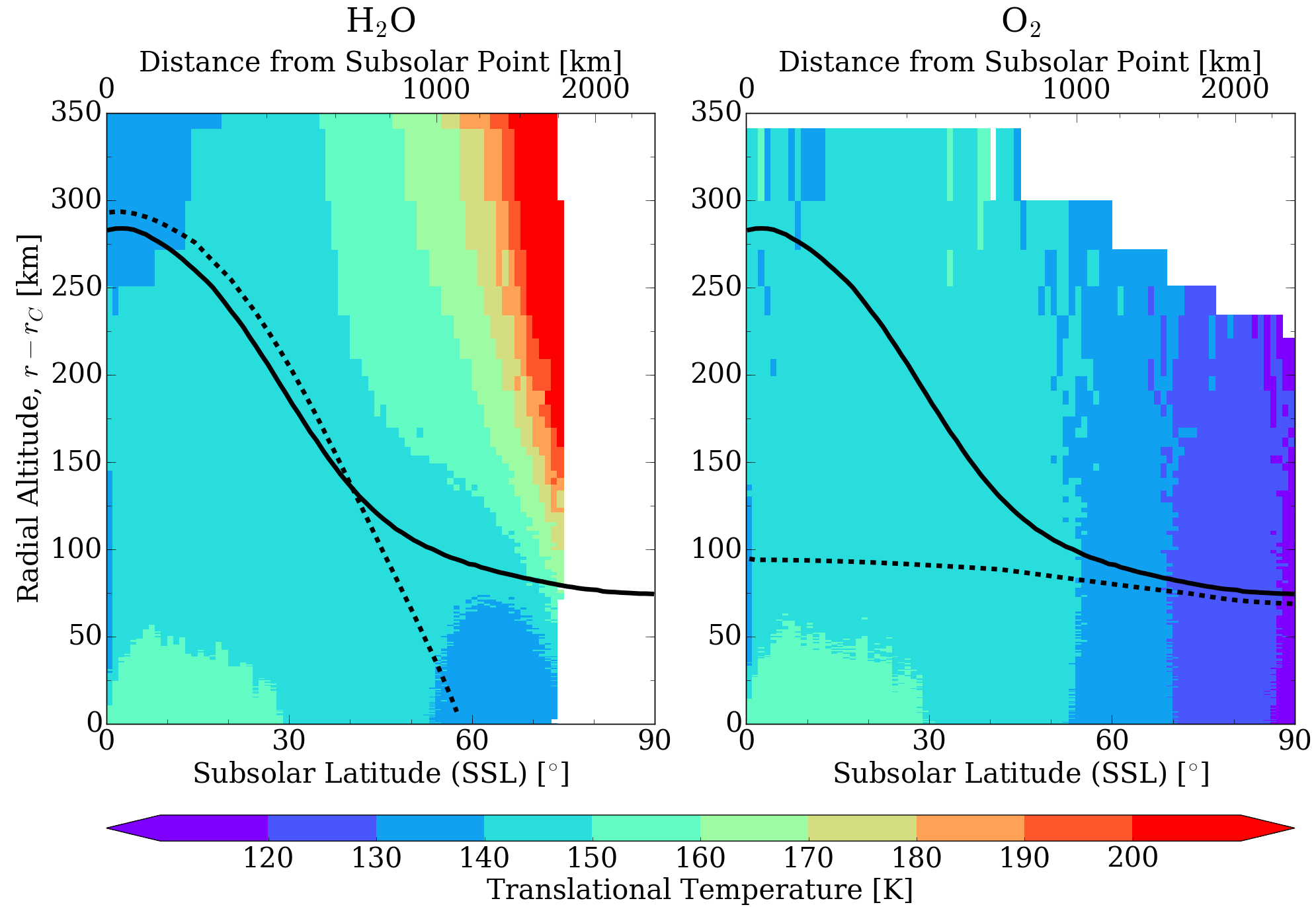}
    \caption{Translational temperature (color spectrum) in a O$_2$+H$_2$O atmosphere vs. subsolar latitude (SSL; x-axis, \textit{bottom}), distance from subsolar point (x-axis, \textit{top}), and radial altitude, $r-r_C$ (y-axis). The average exobase, Kn$_\mathrm{avg}=\frac{\mathrm{Kn}_\mathrm{H_2O} n_\mathrm{H_2O} + \mathrm{Kn}_\mathrm{O_2} n_\mathrm{O_2}}{n_\mathrm{H_2O} + n_\mathrm{O_2}} \sim 1$, is represented by a solid black line and the exobase calculated in H$_2$O and O$_2$ single-species atmospheres is represented by dashed black lines. Note the white regions of the plot represent cells with $<$10 particles. See Fig. \ref{fig:O2H2O_TransTemps_SS} in Appendix \ref{profiles} for a comparison to translational temperatures in single-species H$_2$O and O$_2$ atmospheres.}
    \label{fig:O2H2O_TransTemp}
\end{figure}

Assuming a negligible amount of H$_2$ is present, a collisional O$_2$+H$_2$O atmosphere was simulated. Steady-state for either species was reached in about the same time as in their single-species atmosphere. The individual density structures are only marginally affected by the other species. However, inter-species collisions have a significant effect on the thermal structure of the atmosphere. Although far from the subsolar point (SSL $\gtrsim$45$^\circ$, $\gtrsim$700 km) the H$_2$O component is no longer dense enough for H$_2$O-H$_2$O collisions to occur, it remains in a collisional environment due to the background O$_2$. As a result, as can be seen in Fig. \ref{fig:O2H2O_TransTemp}, the local atmospheric temperature is determined by O$_2$, and the large H$_2$O temperatures generated far from the subsolar point in the single-species atmosphere are reduced by collisions with the O$_2$. Similarly, near the subsolar point the H$_2$O component determines the local atmospheric temperatures. These changes in temperatures affect the local scale heights and the average exobase altitudes of the two species, Kn$_\mathrm{avg}=\frac{\mathrm{Kn}_\mathrm{H_2O} n_\mathrm{H_2O} + \mathrm{Kn}_\mathrm{O_2} n_\mathrm{O_2}}{n_\mathrm{H_2O} + n_\mathrm{O_2}} \sim 1$.

\subsection{H$_2$ Component}

\subsubsection{Single-Species}

Assuming H$_2$ is present at some level, as described earlier we implemented a range of H$_2$ surface fluxes that roughly correspond to the following global surface densities: $n_0 \sim 4 \times 10^8$ cm$^{-3}$ (``upper bound''), which was implemented in C20, $n_0 \sim 4 \times 10^7$ cm$^{-3}$ (``intermediate''), and $n_0 \sim 4 \times 10^6$ cm$^{-3}$ (``lowest value''). At the lowest flux the exobase is at the surface, whereas for the higher values it is located above the surface. However, in all cases the calculated escape rates differ when including or neglecting collisions. The collisional and ballistic H$_2$ atmospheres reached steady-state in $\sim$3--4$ \times 10^5$ seconds and $\sim$1--2$ \times 10^5$ seconds, respectively.

\begin{figure}[h!]
    \centering
    \includegraphics[scale=0.25]{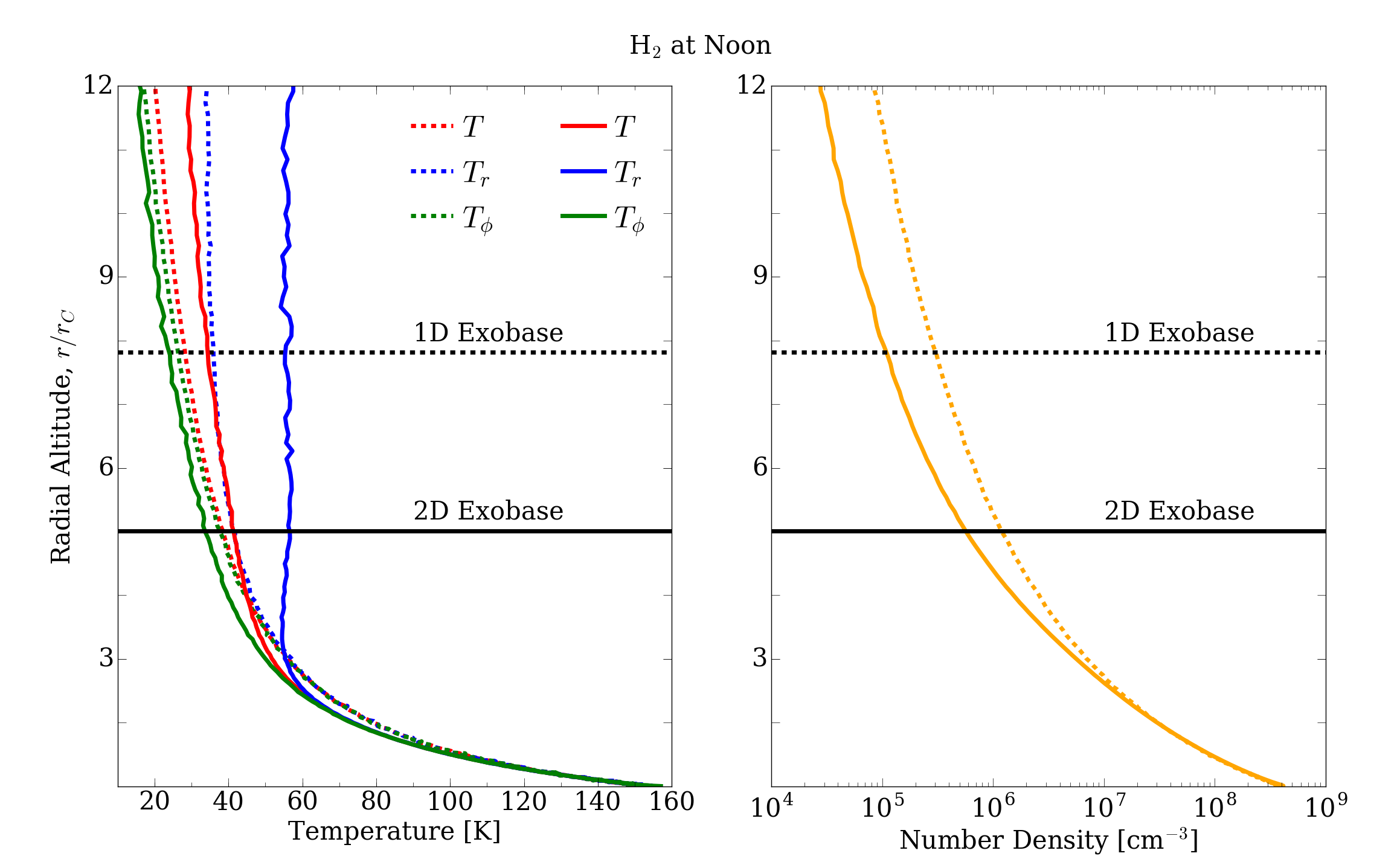}
    \includegraphics[scale=0.25]{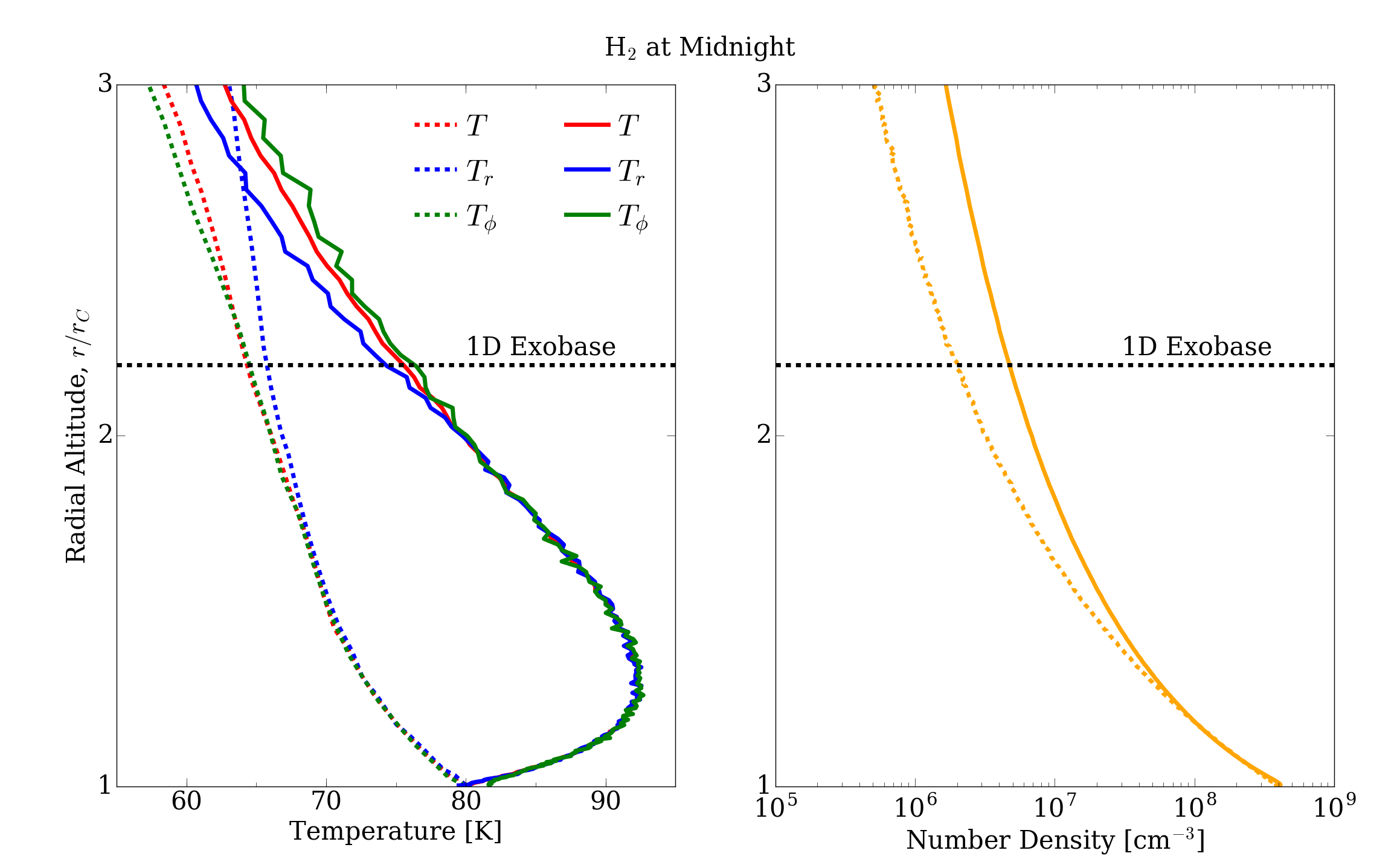}
    \caption{Temperature (\textit{left}) and number density (\textit{right}) profiles vs. radial altitude, $r/r_C$, in a collisional H$_2$ atmosphere. Results from 1D simulations (dashed colored lines) at noon (\textit{top}) and midnight (\textit{bottom}), taken from C20, are compared to those in a 2D simulation (solid colored lines) at SSL=0$^\circ$ and SSL=180$^\circ$, respectively. The nominal exobase (Kn = $\ell_\mathrm{MFP}(r)/H(r) \sim 1$) in the 1D and 2D atmospheres are represented by the dashed and solid black lines, respectively.}
    \label{fig:H2_Temps_Dens_Noon_Midnight}
\end{figure}

Due to the large scale heights and ballistic orbits the H$_2$ experiences extensive transport and escape. Not surprisingly the exobase for the collisional cases will occur at higher altitudes near midnight and lower near noon than in 1D simulations in C20, as indicated in Fig. \ref{fig:H2_Temps_Dens_Noon_Midnight}. Moreover, due to transport, the densities differ in the opposite way, less dense near noon and more dense near midnight in 2D. Such transport can also heat the night-side atmosphere via collisions well below the nominal exobase producing a maximum in the temperature profile above the surface. It is also seen that the various components of the temperature diverge with increasing altitude as the collision rate diminishes starting at about a scale height below the nominal exobase, but in a manner very different from that in a 1D simulation.

This day-to-night transport and the concomitant heating of the night-side atmosphere depend critically on the description of escape which we implement for all H$_2$ particles reaching the Hill sphere. We found that including the effect of Jupiter's gravity and the centrifugal force due to Callisto's orbital motion on the H$_2$ trajectories did not alter the atmospheric structure, but did enhance local escape rate away from and towards Jupiter by a few percent.

The total escape rate in the 2D H$_2$ collisional atmospheres, with a noon-to-midnight temperature gradient is, as expected, less than that of a 1D H$_2$ atmosphere. For the upper bound case, in both the 1D and 2D atmospheres, even with the additional Hill sphere losses, collisions suppress total escape rates relative to those obtained in ballistic simulations and when integrating the analytic Jeans escape rate across the surface (note the escape rates in these ballistic simulations are slightly larger than the integrated Jeans escape rate because of the additional losses to the Hill sphere). This reduction in escape via collisions, however, has only a small effect on the H$_2$ density profiles. Consistent with C20, the thermal escape rates are more than 2 orders of magnitude smaller than the rates at which particles return to the surface.

Ignoring other loss processes, such as photochemical and plasma-induced reactions in the atmosphere, as well as reactions in the regolith, the loss due to escape depends on the unconstrained values of the assumed surface flux. For the upper bound and intermediate cases, the required net production of new H$_2$ is significantly larger than might be expected based on radiolysis of an ice covered surface in Callisto's plasma environment (c.f., \citealt{vorburger2018} and references therein), thereby requiring production of H$_2$ by other sources as discussed earlier. Therefore, better constraints on H$_2$ production are critical. We also note that reduction in the loss rate is not simply linear with reduction in the surface flux. For example, when the surface flux drops by an order of magnitude from the upper bound to the intermediate case, the escape rate only drops off by a factor of $\sim$2 and collisions enhance rather than suppress escape relative to ballistic simulations. This nonlinearity in the change in escape rate with change in the surface flux is consistent with 1D simulations \cite{volkov2011} (e.g., Fig. 10 therein) for similar Jeans parameters. In addition, in the less dense atmosphere with a lower exobase, day-to-night transport is actually more efficient, producing a warmer maximum temperature above the surface near midnight, $\sim$100 K.  Finally, similar to the intermediate case, the escape rate for the lowest value case is about 40$\%$ larger when considering collisions than when they were neglected. As a result, for both the intermediate and lowest value cases, the escape rate is a larger fraction of the average surface source ($\sim$0.03) relative to the upper bound case ($\sim$0.006).

\subsubsection{O$_2$+H$_2$O+H$_2$}

Collisional O$_2$+H$_2$O+H$_2$ atmospheres were simulated. Steady state for each species was reached in about the same time as in their single-species atmospheres. As in C20, for the largest H$_2$ surface flux, it is the dominant collision partner with the heavier species because its scale height is roughly an order of magnitude larger than that of the O$_2$ and H$_2$O components. Indeed, by the altitude H$_2$ density has decreased with altitude by only an order of magnitude, the densities of the heavier species have dropped off several orders of magnitude. Thus, this relatively dense H$_2$ component primarily determines the local atmospheric temperatures throughout the atmosphere, and the large, non-Maxwellian H$_2$O temperatures generated far from the subsolar point of the single-species atmosphere are fully reduced. 

Even when the H$_2$ surface flux is reduced by a factor of 10, although less efficient, it still influences the local temperature throughout the atmosphere because, again, it becomes the most dense atmospheric component within an additional scale height of the other species. Moreover, the rate of cooling with increasing altitude differs for H$_2$ depending on its surface flux and corresponding escape rate, which then affects how much it actually cools the other species. An interesting result is that even at the lowest density the H$_2$ component, which is too thin for H$_2$+H$_2$ collisions to occur frequently, can still cool the other species as it diffuses through the collisional H$_2$O and O$_2$. However, not surprisingly, this occurs primarily at the lower SSL where the atmosphere has the highest collision rate, and the cooling rate with altitude again differs relative to the more dense H$_2$ components because of the difference in escape rates. 

At low SSL, collisions with H$_2$O will actually enhance migration of the H$_2$ to higher SSL, a process more prevalent for the lower density H$_2$ components as illustrated later in Fig. \ref{fig:1dprofiles} when presenting suggested JUICE flyby detection thresholds. It is seen that near the surface, up to $\sim$200 km and $\sim$400 km for the intermediate and lowest value cases, respectively, the local H$_2$ density near noon is less than that at the terminator and midnight until the local H$_2$O density is eventually less than that of the H$_2$; and above which the local H$_2$ density near noon begins to increase relative to the other local densities, until eventually becoming the most dense part of the atmosphere. For the upper bound case, the local H$_2$ density near noon remains the largest among that at the terminator and midnight even when it is less dense than the H$_2$O ($\lesssim$100 km), but when comparing this profile to that of the single-species atmosphere, the local density near noon is larger in the latter as there is no enhanced migration away from this region.

Since H$_2$ only interacts with the other species relatively close to the surface, the H$_2$ densities in the upper atmosphere and escape rates of single- and multi-species H$_2$ atmospheres are largely unaffected. The heating of the night-side atmosphere via the day-to-night H$_2$ transport has little effect on the night-side O$_2$ as the temperature maximum seen in Fig. \ref{fig:H2_Temps_Dens_Noon_Midnight} and the altitude where this occurs in the 10$\times$ less dense H$_2$ atmosphere occurs at an altitude where the O$_2$ density is negligible.

\section{Discussion} \label{Sec4}

The implications of our results, the effects of our assumptions, and the factors not treated in these simulations are discussed below.

\subsection{The Presence of H$_2$}

Based on the surface fluxes used here, in agreement with \cite{roth2017}, sublimated H$_2$O can be a significant source of H from the day side atmosphere. However, even when H$_2$ has a surface density multiple orders of magnitude less than that of H$_2$O, as a result of its large scale height, significant densities occur out to the Hill sphere. Thus, whereas sublimated H$_2$O produces H predominantly in the subsolar region, H$_2$ can be a global source of H. Nascent hot H atoms produced from water vapor near the subsolar point would have to diffuse through the collisional H$_2$O and O$_2$ components close to the surface as well as any H$_2$ that might be present. However, the production of H at high altitudes from H$_2$ readily contributes to the H corona. Thus, although the H$_2$ photodissociation and dissociative ionization rates are much smaller than those of H$_2$O, as shown in Table \ref{tab:reaction_rate}, H$_2$ could be a significant source of the H corona in the terminator region as described below.

\begin{figure}[h!]
    \centering
    \includegraphics[scale=0.28]{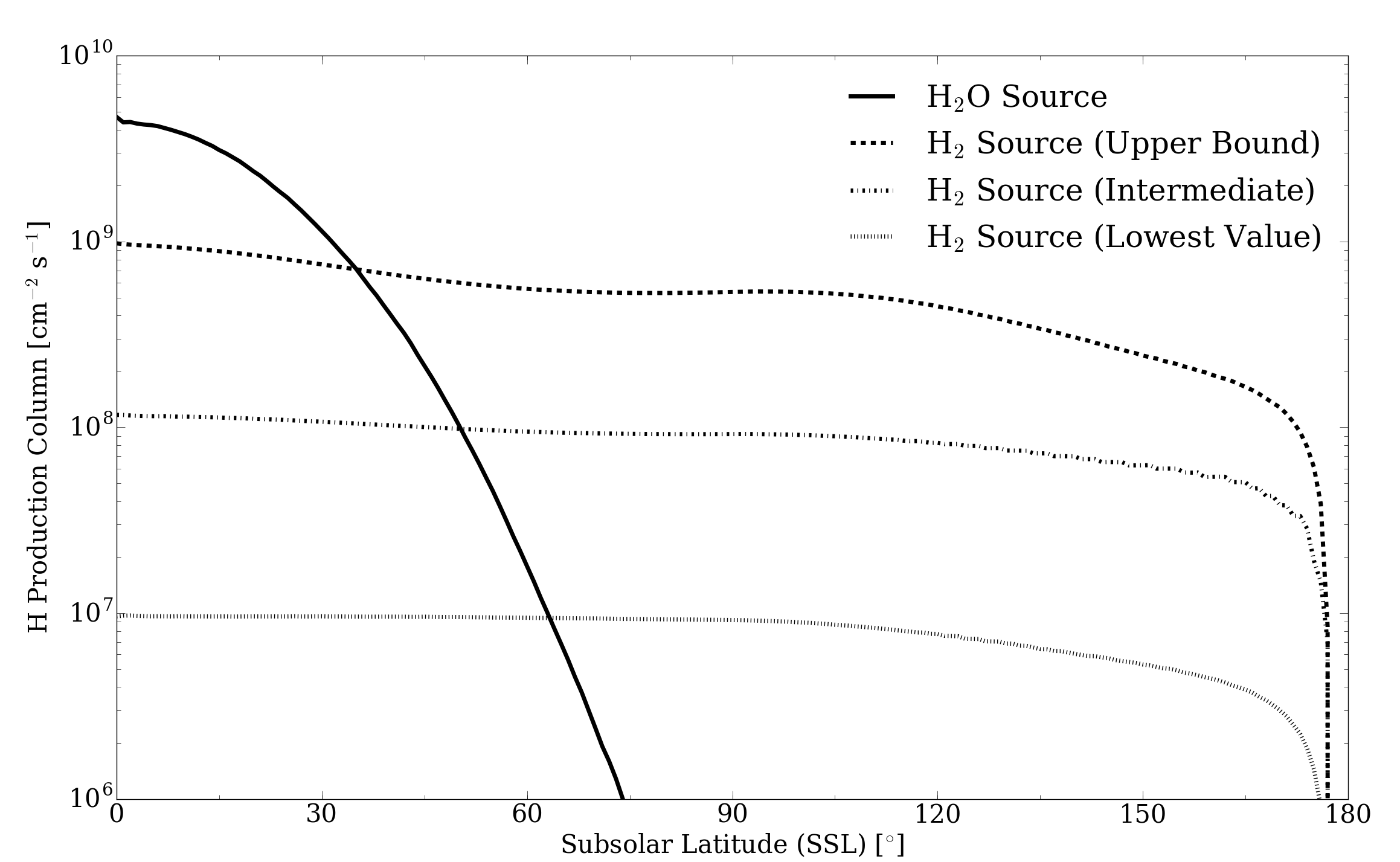}
    \caption{Radial column-integrated H production rate from photodissociation and photodissociative ionization of H$_2$ and H$_2$O in the O$_2$+H$_2$O+H$_2$ atmospheres considered here vs. SSL.}
    \label{fig:H_Production}
\end{figure}

Ignoring possible atmospheric absorption of the incoming solar flux and subsequent interactions of the nascent H with the other atmospheric components, Fig. \ref{fig:H_Production} illustrates the local radial column-integrated production rates of H by H$_2$O and H$_2$ calculated using the density distributions from the O$_2$+H$_2$O+H$_2$ atmospheres considered here with the differing amounts of H$_2$ using the rates in Table \ref{tab:reaction_rate}. With the model surface fluxes used here, H$_2$O produces a relatively large amount of H near the subsolar point, but its production rate diminishes rapidly with increasing altitude and distance from the subsolar point. Conversely, the relatively low production rate of H from H$_2$ occurs globally, except in Callisto's shadow, and diminishes slowly with SSL. Indeed even for the lowest surface flux, where its surface density is $\sim$3 orders of magnitude less than that of the H$_2$O it produces more H beyond SSL $\sim$ 60$^\circ$. Indeed a more global source appears to be suggested by the fact that the observed line-of-sight column density of the H corona was the most pronounced near and beyond the terminator (\citealt{roth2017}; see Fig. 4 therein). These results, of course, scale roughly with the highly unconstrained surface rates for both species. Thus, it is important to constrain the source rates of H$_2$O and H$_2$.

\setlength{\tabcolsep}{10pt}
\renewcommand{\arraystretch}{1.5}
\begin{table}[h!]
\centering
\caption{Production Rate of H via Photodissociation and Photodissociative Ionization}
\begin{tabular}{|l|c|}
\hline
Reaction & Rate$^{a, b}$ \tnote{a, b} [s$^{-1}$]\\
\hline
H$_2$O + h$\nu$ $\rightarrow$ H + OH & 5.16(-7) \\
\hline
H$_2$O + h$\nu$ $\rightarrow$ O + H + H & 4.93(-8) \\
\hline
H$_2$O + h$\nu$ $\rightarrow$ OH$^+$ + H + e & 3.82(-9) \\
\hline
H$_2$ + h$\nu$ $\rightarrow$ H + H & 2.90(-9) \\
\hline
H$_2$ + h$\nu$ $\rightarrow$ H + H (2s, 2p) & 2.15(-9) \\
\hline 
H$_2$ + h$\nu$ $\rightarrow$ H$^+$ + H & 6.92(-10) \\
\hline 
\end{tabular}
\label{tab:reaction_rate}
\begin{tablenotes}\footnotesize
    \item[a] $^a$ Calculated for an ``average'' Sun \citep{huebner2015} and scaled to 5.2 AU ignoring absorption with depth into the atmosphere
    \item[b] $^b$ The number in the parentheses represents powers of 10
\end{tablenotes}
\end{table}

Since any radiolytically produced H$_2$ that escapes from Callisto, but not from Jupiter's field, has a lifetime longer than Callisto's orbital period (Table \ref{tab:lifetimes}),  a neutral H$_2$ torus could form, as is the case at Ganymede \citep{marconi2007} and Europa \citep{smyth2006, smith2019}. Indeed, although the source rate can be much lower, the H$_2$ lifetime is much longer than that of the other icy Galilean satellites, and thus an H$_2$ torus at Callisto might be detectable. Since the torus density roughly scales with the escape rate (e.g., \citealt{johnson1990}), the detection of an H$_2$ torus would help constrain the surface source rate as well as provide insight into local responses to other magnetospheric processes. While direct detection of an optically thin neutral H$_2$ torus would be difficult, toroidal interactions with the local plasma can produce H$_2^+$ ions and/or energetic neutral atom emissions, thereby making indirect detection feasible.

\subsection{JUICE}

\begin{figure}[h!]
    \centering
    \includegraphics[scale=0.28]{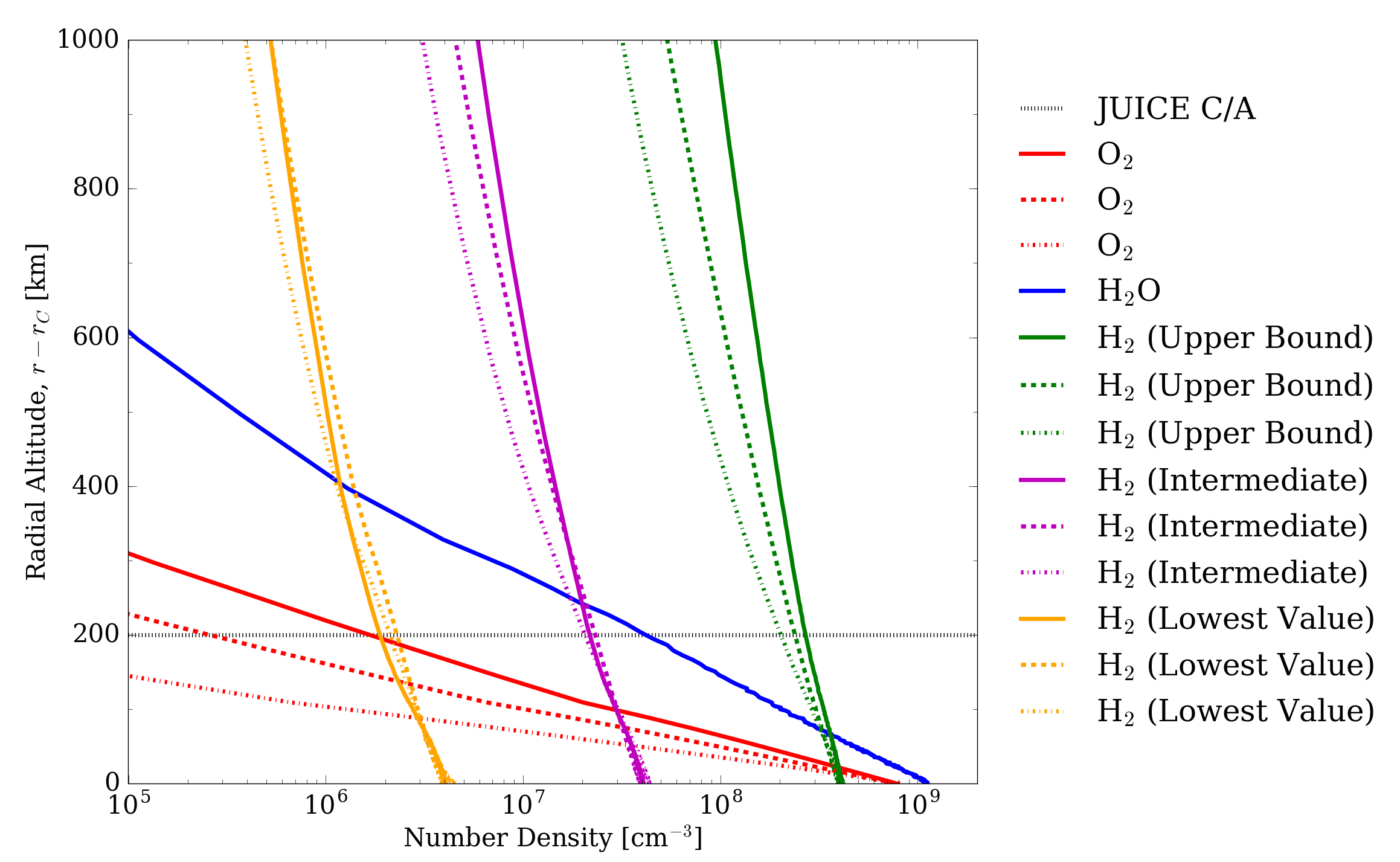}
    \caption{Density profiles of the three neutral species H$_{2}$O (blue line), O$_{2}$ (red lines), and H$_{2}$ (green lines) at the subsolar point (solid lines), the terminator (dashed lines), and the anti-solar point (dash-dotted lines). The dotted black line represents the closest approach (C/A) altitude for JUICE flybys.}
    \label{fig:1dprofiles}
\end{figure}

The next spacecraft expected to visit Callisto and study its atmosphere is JUICE. Before entering into orbit around Ganymede, JUICE will perform a dozen flybys at Callisto at various illumination angles and above the polar and equatorial regions; the closest approach (C/A) will be at 200 km altitude for several of these flybys \citep{boutonnetjuice}. During these flybys, JUICE will measure the neutral gas densities of the icy moon with the Neutral gas and Ion Mass spectrometer (NIM) (\citealt{fohn2021} submitted). Figure \ref{fig:1dprofiles} illustrates the simulated densities of the neutral H$_{2}$O, O$_{2}$, and H$_{2}$ at the subsolar point, the terminator, and the anti-solar point. The sublimated H$_2$O does not reach the latter two regions and the O$_2$ component above 200 km is very tenuous on the night-side. Thus, direct detection of H$_{2}$O by NIM, as well as a thermal O$_{2}$ component, will become more challenging with increasing altitude and distance from the subsolar point. The other particle densities presented will be easily detectable with NIM since the preliminary flight model already at its current stage has a signal-to-noise detection limit of a few 100 cm$^{-3}$ for an integration time of 100 s (\citealt{fohn2021} submitted). Therefore, the direct detection of H$_2$, possible even at the lowest source rate used here, or even a non-detection, would provide a critical constraint on the nature of and source rates in Callisto's atmosphere.

\subsection{Caveats}

The results presented focused on simulations of atmospheres driven only by the surface temperature. This includes a simplified model of the ice sublimation rate and ignores the role of the regolith on the returning molecules as well as those atmospheric processes induced by the solar radiation and the plasma, using only an estimate of the O$_2$ column density as a rough guide for the effect of radiolysis. The photo-processes have the same time scales as those at Ganymede, which have been shown to influence the evolution of its atmosphere (e.g., \citealt{marconi2007, leblanc2017} and will be important for Callisto, especially in producing the detected H corona, but the plasma-induced processes at Callisto are significantly reduced due to the orders of magnitude lower plasma densities as well as by any electrodynamic interactions with the Jovian magnetosphere (e.g., \citealt{strobel2002}).

Since Callisto is tidally locked to Jupiter, its orbital period, $t_\mathrm{orb}$ $\sim$ 1.4$\times10^6$s, and the length of its day are the same. For the assumed surface fluxes Table \ref{tab:lifetimes} gives the approximate time it takes each species to reach steady-state, as well as the lower limits to relevant lifetimes are normalized by $t_\mathrm{orb}$. As can be seen, even when the initial state of the atmosphere is a vacuum, steady-state is reached for each species in a fraction of the time it takes Callisto to complete 1 orbit, suggesting 2D steady-state simulations which assume Callisto is stationary in orbit can be used to estimate the local conditions for an evolving unsteady atmosphere. It is also seen that steady-state is reached very rapidly relative to the photo- and plasma-induced losses. Therefore, although the day-to-night transport of H$_2$ in the exosphere and the concomitant escape rates will likely differ when such processes are considered, the principal structure will not differ significantly. Note, we did not include lifetimes induced by interactions with the ionosphere, which has been suggested to be a transient phenomena (e.g., \citealt{kliore2002}), nor have we considered the atmospheric evolution on Callisto during which the surface fluxes can change as the atmosphere evolves.

\begin{table}[h!]
    \centering
    \caption{Lifetimes}
    \begin{tabular}{|c|c|c|c|c|c|}
        \hline
        \multirow{3}{*}{Species} & \multicolumn{5}{c|}{Times / $t_\mathrm{orb}$ $^a$ \tnote{a}} \\
        \cline{2-6}
                & Steady    & Photo$^b$ \tnote{b}    & Photo$^b$    & Plasma$^c$ \tnote{c}   & Plasma$^c$ \\
                & State     & ioni.     & diss.     & ioni.     & diss. \\
        \hline
        H$_2$O  & 0.01      & 32        & 1.3       & 45        & 38 \\
        \hline
        O$_2$   & 0.02      & 23        & 3.6       & 42        & 240  \\
        \hline 
        H$_2$   & 0.25      & 220       & 240       & 62        & 150  \\
        \hline
    \end{tabular}
    \label{tab:lifetimes}
    \begin{tablenotes}\footnotesize
    \item[a] $^a$ Callisto's orbital period and length of its day, $t_\mathrm{orb} \sim 1.4 \times 10^6$ s
    \item[b] $^b$ Photo. lifetimes are calculated for an ``average'' Sun \citep{huebner2015} and scaled to 5.2 AU ignoring possible absorption with depth into the atmosphere
    \item[c] $^c$ Based on plasma properties from \cite{vorburger2019} and estimates for plasma interaction cross-sections from \cite{Tawara1990, Straub1996, Deutsch2000, Luna2005, Riahi2006, McConkey2008}
    \end{tablenotes}
\end{table}

Since steady state is achieved for the neutrals in the collisional regime in times much shorter than the orbital period and many orders of magnitude shorter than lifetimes induced by interactions with solar photons and the co-rotating Jovian thermal plasma, this work can be used to guide our understanding of the principal components of the atmosphere. Non-thermal interactions will certainly affect the escape rates and can produce detectable emissions. In addition, the varying illumination of Callisto's surface and atmosphere as it orbits Jupiter will affect local production and depletion rates.

Assumptions about the role of how the dark lag material that dominates the surface of the very porous regolith can affect the H$_2$O sublimation rate as well as the production of H$_2$ and O$_2$ volatiles need to be examined. A critical question is the role of porosity of and reactions in Callisto's regolith on the returning volatiles. Since O$_2$ is more reactive in the regolith and more readily dissociated in the gas phase and H$_2$ can escape, constraining the relative, steady state surface fluxes is difficult. The role of thermal inertia on the sublimation of the water vapor needs to be examined, as at the other icy Galilean satellites (e.g., by \citealt {leblanc2017} at Ganymede and \citealt{oza2019} at Europa), and the influence of local cold trapping (e.g., \citealt{spencer1987b}) need to be addressed. We tested the effects of Callisto's orbital motion and Jupiter's gravitation and found that they did not significantly affect the primary structure. However, the Coriolis force due to the orbital motion would affect our assumption of azimuthal symmetry, and would require more realistic local upper boundaries than simply setting the symmetric Hill sphere as the global upper boundary. The factors discussed above will be considered in future work.

\section{Conclusion} \label{Sec5}

Although Callisto's thin atmosphere is very poorly constrained, it is critical to understand how the interactions between those species thought to be present determine the atmospheric structure. This was explored here by simulating atmospheres with surface densities for O$_2$ and H$_2$O consistent with values suggested in the literature. Although H$_2$ is expected to be present, there is no observational constraint, and a range of densities was examined from no H$_2$ to the likely largest possible value. Therefore, we presented results of 2D DSMC simulations of model atmospheres on Callisto containing sublimated water vapor and radiolytic products, O$_2$ and H$_2$, focusing on their response to the variation of the surface temperature with subsolar latitude. Although the structure of the single component, 2D collisional atmospheres were shown to not differ significantly from ballistic simulations, that is not the case for the multi-component atmospheres consistent with our 1D results in C20. The thermal structure is significantly influenced by the molecular interactions, as well as by the diurnal variation in the surface temperature, and, when H$_2$ is present, thermal escape and concomitant cooling. Since sublimation of water vapor is extremely sensitive to the surface temperature, the density decreases by several orders of magnitude with distance from the subsolar point and the flow transitions from collisional to ballistic in the absence of a background component. With the addition of a global O$_2$ component, the H$_2$O remains collisional out to the terminator and the local exobase is enhanced. The local atmospheric temperatures are determined by the more dense H$_2$O near the subsolar point and transition to being determined by O$_2$ with distance from the subsolar point. When H$_2$ is introduced it becomes an important or dominant collision partner affecting the local atmospheric temperatures. However, since it primarily interacts with the heavier species within $\lesssim$1000 km of the surface, flow in the upper atmosphere and escape rates do not differ significantly from the single-species 2D H$_2$ atmosphere. The extensive day-to-night transport of H$_2$ generates a roughly uniform exobase and heating of the night-side atmosphere. Although highly unconstrained, H$_2$ is produced stochiometrically with O$_2$, which is suggested to be present in Callisto's atmosphere. Since its spatial distribution differs enormously from that of H$_2$O, even at low densities, it could be a significant source of the observed H corona \cite{roth2017} near and beyond the terminator region. The results presented can be used as a guide to proposed telescopic and spacecraft observations while a self-consistent 3D model is developed to account for Callisto's orbital motion.

%We are currently developing a self-consistent 3D model to correctly include the asymmetric, unsteady, and non-thermal processes described above. Nevertheless, the 2D model used here, which is a crucial step towards a more holistic 3D model, is a significant improvement on our previous 1D model, as we implemented a temperature gradient across the surface and simulated the corresponding local transport phenomena. Moreover, the results presented can be used as a guide to proposed telescopic and spacecraft observations. Indeed the integration of models and observations are needed to determine the nature of the atmosphere on this unique body.

\newpage
\appendix
\counterwithin{figure}{section}

\section*{Appendix}

\section{Parallel Implementation} \label{parallel}

The DSMC method can be used for a wide variety of fluid problems and is implemented in a variety of ways. Our New York University implementation is designed for flexibility; it can run in single-threaded mode on laptops, in multi-threaded mode (OpenMP) on laptops and computer servers, as well as in single program multiple data (SPMD) mode \citep{darema2011} using MPI \citep{walker1992, gropp1999} on large-scale computers. Relatively small 1D problems, such as those carried out in C20, can be solved with OpenMP's fine-grain parallelism, while large 1D and 2D problems, such as the simulations of this study, benefit from the scalability of SPMD coarse-grain parallelism.

The domain decomposition method used is also designed for flexibility. For instance, a 2D problem can first be split along the radial altitude axis, and then further split along the angular axis, resulting in a 2D domain decomposition. A 2D grid problem can also be split along the radial altitude or angular axis alone, resulting in a 1D domain decomposition. The choice can be based, for example, on the number of particles in the simulation, the complexity of their interactions, and the processing efficiency sought. As a result of domain decomposition, a problem space is divided in a set of sub-domains, where each sub-domain is assigned to an MPI rank (a process executing on one or more processor core).

Decomposing a problem space into sub-domains can then be computed in parallel (e.g., \citealt{dietrich1996, lebeau1999}), where each sub-domain encompasses (a) distributing particles either during the initialization phase or as an on-going process at regular simulated time intervals, (b) calculating particle movement and collisions, and (c) transferring particles whose updated trajectories have positioned them in a different sub-domain. As described in Section \ref{2Dmodel}, in the present case particles are injected into an initially empty domain. The transfer of particles from one sub-domain to another requires cooperation between MPI ranks. An asynchronous Halo communication mechanism \citep{fukazawa2016} between neighboring MPI ranks is implemented, where at the end of each time-step all MPI ranks exchange a list of particles that need to move to a new sub-domain.

The molecular kinetics simulation results relating to Callisto's atmosphere we present herein were obtained with the DSMC method using 1D and 2D grids; the former are divided in equal radial sub-domains, while the latter are divided in equal angular sub-domains, each covering the entire radial altitude axis. Thus, the 2D grids are processed using a 1D domain decomposition method using MPI.

\section{2D Grid} \label{2dgrid}

\begin{figure}
    \centering
    \includegraphics[scale=0.22]{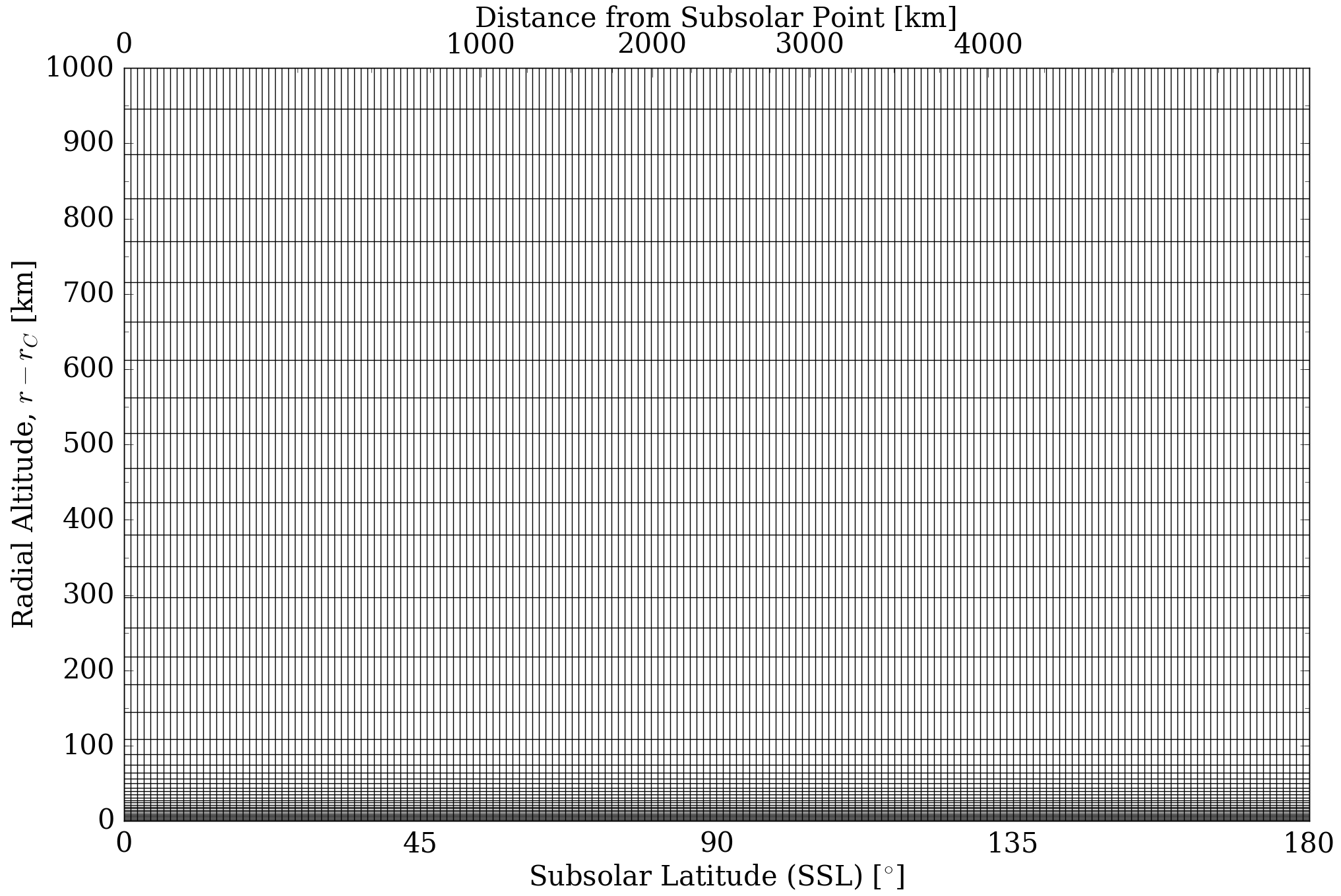}
    \caption{Example of a 2D computational grid up to $r-r_C = 1,000$ km (y-axis) from subsolar latitude (SSL)$=0^\circ \rightarrow 180^\circ$ (x-axis, \textit{bottom}) and with increasing distance from the subsolar point (x-axis, \textit{top}) with non-uniform radial cells and uniform angular cells.}
    \label{fig:2D_Grid}
\end{figure}

Assuming an isothermal atmosphere set to the approximate noon surface temperature, $T_0 = T_\mathrm{noon}$, we calculate the radial extent of each cell as equal to the local mean free path (MFP), $\ell_\mathrm{MFP}(r, T_0)$, where $r$ is the radial altitude. In a multi-species simulation, the radial extents of cells are determined according to the average $\ell_\mathrm{MFP}(r, T_0)$ among the species: $\ell_{\mathrm{MFP}, \mathrm{avg}}(r, T_0) = \frac{\sum_i \ell_{\mathrm{MFP},i}(r, T_0) n_i(r, T_0)}{\sum_i n_i(r, T_0)}$, where $i$ represents the species. As a result, in a multi-component atmosphere with O$_2$, H$_2$O, and H$_2$, the radial extent of cells are represented by the heavier, more dense components near the surface and by the extended H$_2$ component at higher altitudes. In the rarefied regime of single- and multi-component atmospheres comprised of O$_2$ and H$_2$O, $\ell_\mathrm{MFP}(r, T_0)$ or $\ell_{\mathrm{MFP}, \mathrm{avg}}(r, T_0)$ can grow to the order of the planetary body's radius. In such instances, the local (or, in the case of a multi-species atmosphere, average local) atmospheric scale height, $H(r) = \frac{k_B T_0 r^2}{G M m}$ (or $H_\mathrm{avg}(r)= \frac{\sum_i H_i(r) n_i(r, T_0)}{\sum_i n_i(r, T_0)}$), is used instead to define the radial extent of a cell. Here $G$ = 6.674$ \times 10^{-11}$ m$^3$ kg$^{-1}$ s$^{-2}$ is the gravitational constant, $m$ is the mass of the atmospheric species, and $k_B$ = 1.38$ \times 10^{-23}$ m$^2$ kg s$^{-2}$ K$^{-1}$ is the Boltzmann constant. For cell widths along the angular axis that exceed $\ell_\mathrm{MFP}(r)$, we implement sub-cells by dividing the extent of a cell along the angular axis by $\ell_\mathrm{MFP}(r)$. Finally, an example of a 2D grid with non-uniform radial cells and uniform angular cells can be seen in Fig. \ref{fig:2D_Grid}.

\section{Particle Weights} \label{partweight}

The weight of each particle, $W_P$, is determined via the following equation:

\begin{equation}
    W_P = \frac{ \Phi_0 A \Delta t}{N_\mathrm{inj}}.
\end{equation} \label{PartWeight}

\noindent Here $\Phi_0 = \frac{1}{4} n_0 v_{th}$ is the local molecular flux at the surface, $A$ is the surface area of the cell the particle is being injected into, and $N_\mathrm{inj}$ is the number of particles injected into the cell at each time-step. For cells at the surface, $A = 2 \pi r_C^2 \int_\mathrm{SSL1}^\mathrm{SSL2} \cos$ (SSL) $d$SSL, where SSL1 and SSL2 are the lower and upper SSL bounds of the cell, respectively, and the azimuth axis wraps around the axis of symmetry from $0 \rightarrow 2 \pi$. With constant $\Delta$SSL, $A$ can vary significantly across the surface; e.g., at the axis of symmetry $A$ is $\sim$2 orders of magnitude smaller than at the terminator.

For O$_2$ and H$_2$ particles $N_\mathrm{inj}$ is kept constant for each surface cell so that $W_P$ varies according to the local molecular flow rate ($\Phi_0 A$) but is fixed in radial distance from the surface. If a particle crosses into a region with a different weight, then either some fraction of its weight is discarded or it is replicated a certain number of times according to the acceptance-rejection method (e.g., \citealt{combi1996}). For the former case, if a particle from one weighted region crosses into a region with a larger weight attributed to it, it successfully enters the domain with the new region's weight if a randomly generated number between 0 and 1 is less than the ratio of the two regions' weights, otherwise it is discarded. Conversely, for the latter case, if a particle from one weighted region crosses into a region with a smaller weight attributed to it, it is replicated into at least as many particles as the integer value of the ratio of the two regions' weights, and an additional particle is replicated if a randomly generated number between 0 and 1 is less than the remainder of the ratio. 

This discarding/replicating procedure serves as a means to improve statistics in otherwise particle-poor regions, such as in cells near the axis of symmetry, and generates a more balanced workload distribution for parallel processing. Since O$_2$ and H$_2$ have a uniform surface density, $W_P$ varies between cells along the SSL at most by a factor of $\sim$3 at the axis of symmetry and roughly a factor of 1 everywhere else. Due to the variation of $n_{0, \mathrm{H_2O}}$ along Callisto's SSL, however, the discarding/replicating procedure is not applicable. That is, the exponential decrease in density from noon to midnight eventually exceeds the gradual increase of the surface area from noon to the terminator and the subsequent gradual decrease from the terminator to midnight. As a result, a cascade of replicated H$_2$O particles can occur if a particle originating near noon with a relatively large $W_P$ traverses across multiple weighted regions, where $W_P$ continuously decreases and, thus, the number of replicated particles continuously increases. This cascade of replicated particles generates a severe workload imbalance and can effectively halt a simulation. Therefore, for H$_2$O particles a constant $W_P$ is used and each surface cell has their own local particle injection number, which is determined by the ratio of the local to the maximum molecular flow rate multiplied by a predefined, fixed $N_\mathrm{inj}$. 

At each time-step, each surface cell will inject at least as many particles as the integer result of this product into the domain, and an additional particle is injected according to the acceptance-rejection method described earlier for the remainder. However, because of the steep drop off in $n_{0, \mathrm{H_2O}}$ and hence local molecular flow rate, depending on the original size of $N_\mathrm{inj}$, surface cells at higher SSLs might not inject any particles. In addition, in the simulations of this study H$_2$O particles rarely migrate beyond the terminator. Thus, because of the rapid decline in the source rate with SSL, the H$_2$O atmosphere is truncated to the terminator (SSL$= 90^\circ$) and the relatively small number of particles that cross the terminator are discarded.

Finally, when calculating inter-species collisions the differences in $W_P$ for each species must be accounted for. For example, if a lighter particle collides with a heavier particle, then only the number of atoms or molecules the lighter particle represents would actually be affected by a collision for the heavier particle. If the heavier particles were affected by each collision with a lighter particle, then too much momentum would be transferred and momentum and energy would not be conserved. Thus, we implement the varying particle weight collision technique described by \cite{miller1994} which conserves energy on average. That is, the lighter particle is always affected by a collision with a heavier particle and the heavier particle is only affected according to the acceptance-rejection method for the ratio of the two particle weights.

\section{Collision Parameters} \label{collparams}

Collisions are described using the LB and VHS models with the parameters listed for each species in Table \ref{tab:parameters}. Here we use a molecular diameter of $d$ = 6.2$ \times 10^{-10}$ m for H$_2$O, which \cite{jansen2010} state was able to reproduce known viscosity values for water vapor over a range of temperatures. The same or similar values for $d$ were found in several other references. However, this value results in a larger H$_2$O-H$_2$O reference collision cross-section than has been used in other atmospheres simulated with DSMC where water vapor is present (e.g., \citealt{marconi2007, leblanc2017}). More detailed cross sections could be used as such data become readily available.

\setlength{\tabcolsep}{10pt}
\renewcommand{\arraystretch}{1.5}
\begin{table}[h!]
\centering
\caption{Collision Parameters}
\begin{tabular}{|c|c|c|c|}
\hline
\multirow{2}{*}{Parameters} & \multicolumn{3}{c|}{Species} \\ \cline{2-4}
 & \textbf{H$_2$O$^a$ \tnote{a}} & \textbf{O$_2^b$ \tnote{b}} & \textbf{H$_2^b$} \\
\hline
Internal Degrees of Freedom & 3 & 2 & 2 \\
\hline
VHS Molecular Diameter, $d$ [($\times 10^{-10}$) m] & 6.20 & 4.07 & 2.92 \\
\hline
VHS Viscosity Index, $\omega$ & 1.00 & 0.77 & 0.67 \\
\hline
\end{tabular}
\label{tab:parameters}
\begin{tablenotes}\footnotesize
\item[a] $^a$ Based on data from \cite{jansen2010}
\item[b] $^b$ Based on data from \cite{bird1994}
\end{tablenotes}
\end{table}

\section{Density and Temperature Profiles} \label{profiles}

\begin{figure}
    \centering
    \includegraphics[scale=0.25]{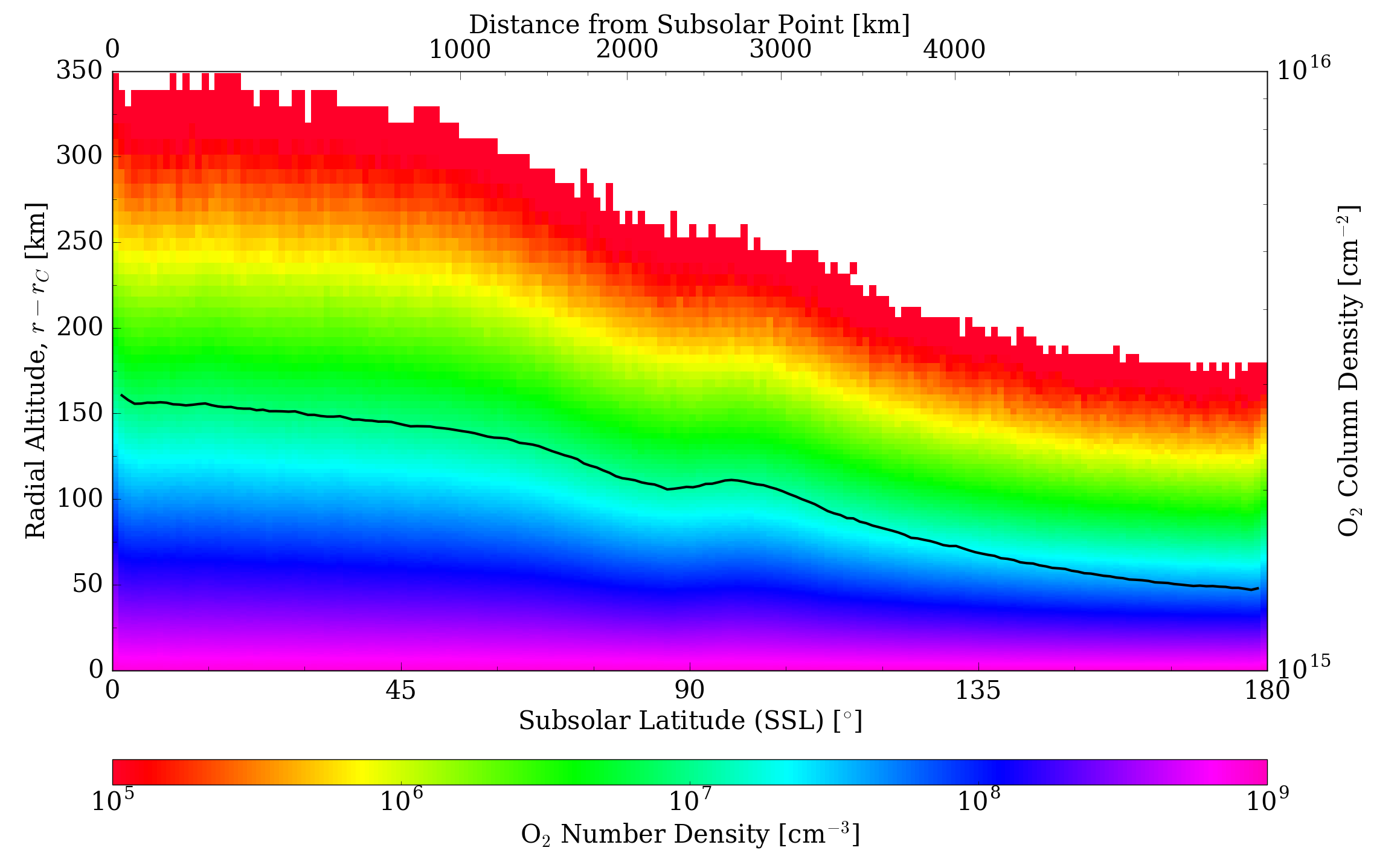}
    \caption{O$_2$ number density (color spectrum) and column density (black line) profiles vs. subsolar latitude (SSL; x-axis, \textit{bottom}), distance from subsolar point (x-axis, \textit{top}), and radial altitude, $r-r_C$ (y-axis).}    
    \label{fig:NumDens_O2}
\end{figure}

\begin{figure}
    \centering
    \includegraphics[scale=0.25]{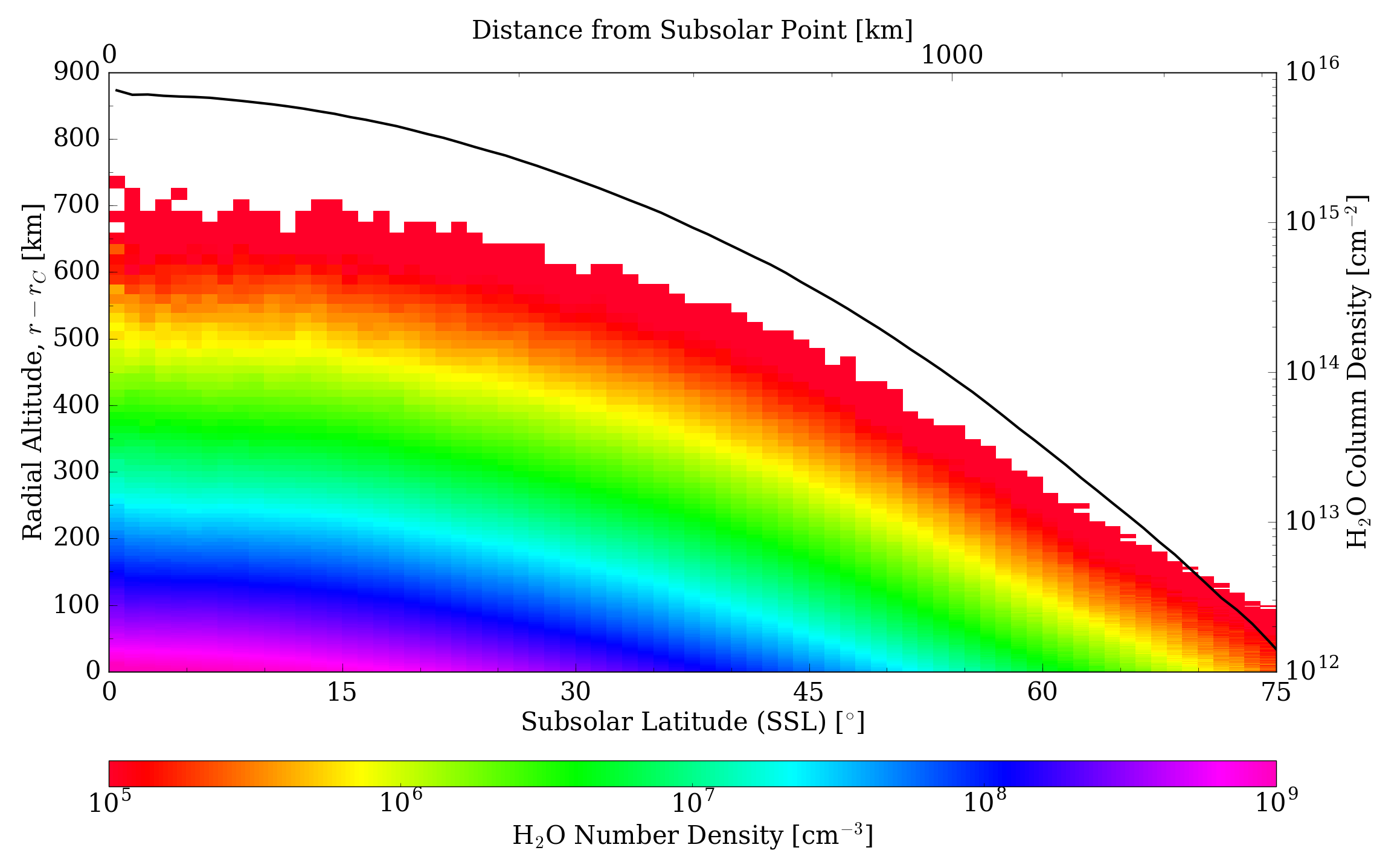}
    \caption{H$_2$O number density (color spectrum) and column density (black line) profiles vs. subsolar latitude (SSL; x-axis, \textit{bottom}), distance from subsolar point (x-axis, \textit{top}), and radial altitude, $r-r_C$ (y-axis).}    
    \label{fig:NumDens_H2O}
\end{figure}

\begin{figure}
    \centering
    \includegraphics[scale=0.22]{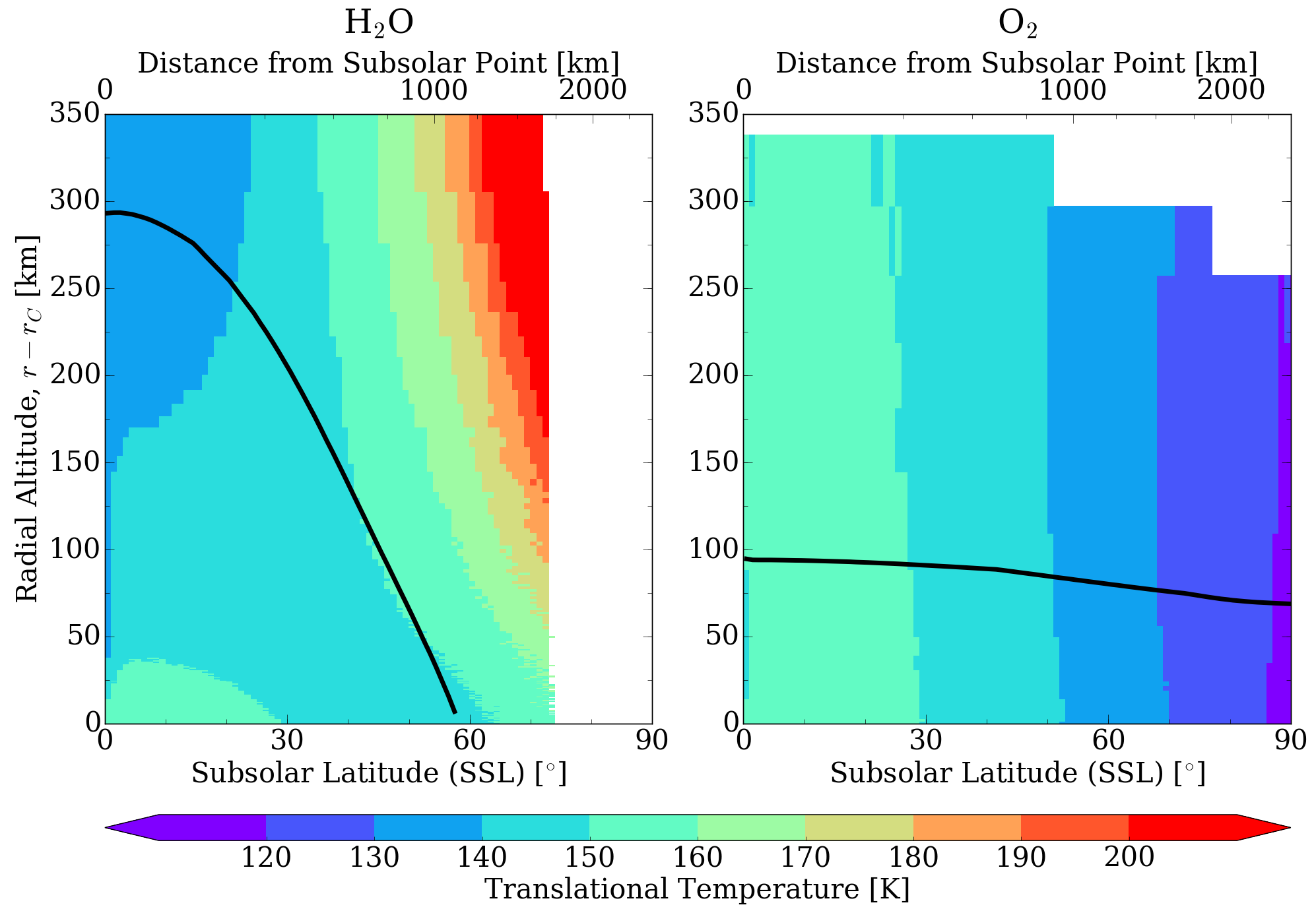}
    \caption{Translational temperature (color spectrum) in single-species H$_2$O (\textit{left}) and O$_2$ (\textit{right}) atmospheres vs. subsolar latitude (SSL; x-axis, \textit{bottom}), distance from subsolar point (x-axis, \textit{top}), and radial altitude, $r-r_C$ (y-axis). The exobase for each atmosphere is represented by a solid black line.}
    \label{fig:O2H2O_TransTemps_SS}
\end{figure}

Number density profiles for O$_2$ and H$_2$O are presented in Figs. \ref{fig:NumDens_O2} and \ref{fig:NumDens_H2O}. Since the density profiles for each species do not significantly vary between single- and multi-species atmospheres, these are taken from the O$_2$+H$_2$O+H$_2$ atmosphere where $n_{0, \mathrm{H_2}} \sim 4 \times 10^8$ cm$^{-3}$ since it has the highest resolution (smallest radial cell widths, most number of cells) among the simulated atmospheres. Translational temperature profiles for H$_2$O and O$_2$ in single-species atmospheres are presented in Fig. \ref{fig:O2H2O_TransTemps_SS} for comparison to those in the multi-species O$_2$+H$_2$O atmosphere in Fig. \ref{fig:O2H2O_TransTemp}.

\newpage
\section*{Acknowledgments}

This work is supported by grants 80NSSC19M0073 and 80NSSC20M0193 from NASA Goddard Space Flight Center’s Solar System Exploration Division. SRCM acknowledges additional support from the NYU Abu Dhabi Global PhD Student Fellowship. The authors also acknowledge M. Marconi for providing invaluable insight when initially developing the models presented here, as well as the two anonymous reviewers whose comments and suggestions greatly improved this manuscript. This research was carried out on the High Performance Computing resources at New York University Abu Dhabi.

\bibliography{CarberryMoganEtAl_2021.bib}

\end{document}